\documentclass[12pt]{article}
\usepackage{graphicx}
\usepackage{float}
\usepackage{subfig}
\usepackage{amsmath}
\usepackage{amsfonts}
\usepackage{tikz}
\usepackage{multirow}
\usepackage{array}
\usepackage{amssymb}
\usepackage{booktabs}
\usepackage{lscape}
\usepackage[authoryear]{natbib}
\usepackage{color}
\usepackage{hyperref}
\usepackage{breakurl}
\usepackage{rotating}
\usepackage{enumitem}
\usepackage{MnSymbol}
\usepackage{comment}
\usepackage{hhline}

\setlist[itemize]{leftmargin=*}
\newtheorem{claim}{{\bf \sc Claim}}
\newtheorem{theorem}{{\bf \sc Theorem}}
\newtheorem{lemma}{{\bf \sc Lemma}}

\newtheorem{proposition}{{\bf \sc Proposition}}

\newtheorem{axiom}{{\bf \sc Axiom}}

\def \Re{\mathbb{R}}
\def \eproof{\hbox{\hskip3pt\vrule width4pt height8pt depth1.5pt}}

\begin{document}

\title{\textbf{Centrality Measures in Networks}\thanks{%
We benefitted from comments by  Gabrielle Demange, Jean-Jacques Herings, Arturo Marquez Flores, Alex Teytelboym, and seminar
participants in different institutions.  We
gratefully acknowledge financial support from the NSF grants
SES-0961481, SES-1155302, SES-2018554, ARO MURI Award No. W911NF-12-1-0509, and from grant FA9550-12-1-0411
from the AFOSR and DARPA} }
\author{Francis Bloch\thanks{%
Universit\'{e}
Paris 1 and Paris School of Economics, francis.bloch@univ-paris1.fr.} \quad
Matthew O. Jackson \thanks{%
Department of Economics, Stanford University, external faculty of the Santa Fe Institute, and fellow of CIFAR, jacksonm@stanford.edu} \quad Pietro Tebaldi \thanks{%
Stanford University, ptebaldi@stanford.edu }}
\date{January 2021}
\maketitle

\begin{abstract}
We show that prominent centrality measures in network analysis are all based on additively separable and linear treatments of
statistics that capture a node's position in the network.
This enables us to provide a taxonomy of centrality measures that distills them
to varying on two dimensions:   (i) which information they make use of
about nodes' positions, and (ii)  how that information is weighted as a function of distance from
the node in question.    The three sorts of information about nodes' positions that are usually used -- which we refer to as ``nodal statistics'' -- are the paths from a given node to other nodes,  the walks from a given node to other nodes,
and the geodesics between other nodes that include a given node.
Using such statistics on nodes' positions, we also characterize the types of trees such that centrality measures all agree, and
we also discuss the properties that identify some path-based centrality measures.

\textsc{JEL Classification Codes:} D85, D13, L14, O12, Z13, C65

\textsc{Keywords:}  Centrality, prestige, power, influence, networks, social networks, rankings, centrality measures
\end{abstract}

\setcounter{page}{0}\thispagestyle{empty}

\newpage
\section{Introduction}

The positions of individuals in a network drive a wide range of behaviors,
from decisions concerning education and human capital \citep*{hahnetal2015} to
 the identification of banks that are too-connected-to-fail \citep{gofman2015}.
Most importantly, there are many different ways to capture a person's centrality, power, prestige, or influence.  As such concepts depend heavily on context,
which measure is most appropriate may vary with the application.  Betweenness centrality is instrumental in explaining the rise of the Medici \citep{padgetta1993,jackson2008,jackson2019},
while Katz-Bonacich centrality is critical in understanding social multipliers in interactions with complementarities \citep*{ballestercz2006},  diffusion centrality is important in understanding many diffusion processes and people's information \citep*{banerjeecdj2013,banerjeecdj2019}, eigenvector centrality determines whether a society correctly aggregates information \citep{golubj2010}, and degree centrality helps us to understand systematic biases in social norms \citep{jackson2016} and who is first hit in
a contagion \citep{christakisf2010}.\footnote{For more discussion and references on the distinction between various forms of influence and social capital see \cite{jackson2020}.}

Despite the importance of network position in many settings, and the diversity of
measures that have been proposed to capture different facets of
centrality, little is known about how to
distinguish the measures.
We provide such a taxonomy by showing that all prominent measures can be viewed as operating in a similar manner, but differing on two dimensions.

In particular,  we  first identify what we call
``nodal statistics.''   These summarize information about the position of a node in the network.
A most basic nodal statistic is the neighborhood structure surrounding a given node: how many nodes are
at various distances from the given node.
Another common nodal statistic is how many nodes lie on walks of various distances from a given node.
Walks allow for cycles and thus result in different information about a node's reach.  Which of paths (neighborhoods) or walks is appropriate depends on the context.
The third most common nodal statistic is a count of how many shortest paths of various lengths between pairs of other nodes a given node lies upon.

Second, we show that each prominent centrality measure can be viewed as
an additively separable weighted average of some nodal statistic.
As weights are varied, the relative importance of a node's position in terms of nearby versus far-off relationships is changed.
Again, which is most natural depends on the context.
In fact, one of our results shows that the most prominent centrality measures are characterized by axioms of monotonicity (higher statistics lead to higher centrality), anonymity (nodes' centralities only depend on their statistics and not their labels), and additivity (statistics are processed in an additively separable manner).   Monotonicity and anonymity are normatively appealing axioms.  Additivity ties things down to a particular functional form, but one that has always been implicitly used when defining centrality measures.  It has
nice mathematical properties, which might explain why it underlies essentially all measures.

From our results, distinguishing centrality
measures boils down to which nodal statistics they incorporate and how they weight information from different distances.
These results help us to categorize most standard centrality measures as being of seven different types, depending on which of the three nodal statistics they make use of and which of four different weighting approaches are used (with some combinations not making sense or being redundant, reducing the twelve to seven).

With this perspective, we provide a discussion of how to interpret which centrality measure is most appropriate in which context.

After having characterized the base way in which centrality measures process information and depend on nodal statistics, we then provide two results that classify specific centrality measures
and the nodal statistics that they use.  These are degree centrality, which only pays attention to immediate connections; and a centrality measure that depends on a weighted neighborhood nodal statistic.
The latter type of centrality measure is tied down by axioms that require that a centrality measure: depend only on shortest paths and not on cycles;  give higher
centrality to nodes that are closer to other nodes closer (all else held constant); and give the same
marginal credit to adding a node at a given distance in different networks.
These theorems are a first step in a broader research agenda to develop characterizations of the many other usual centrality measures and nodal statistics.

The introduction of nodal statistics also helps in identifying social networks for which all centrality rankings coincide.
Because the ranking induced by different centrality measures can be deduced by understanding how their corresponding nodal statistics differ (presuming that weightings are decreasing in distance), centrality measures coincide if and only if all nodal statistics generate the same order. This enables us to identify the trees for which all centrality measures coincide based on an examination of nodal statistics.\footnote{
This is somewhat reminiscent of \citet*{koenigtz2014} who show that many centrality rankings coincide in nested-split graphs, which have a strong hierarchical form.  Trees admit more variation,
and so the characterization here provides new insight,  especially as it helps us understand when nodal statistics coincide.}
In an appendix, we compare network statistics on some simulated networks.

We discuss the associated literature
later in the paper, as it becomes relevant.
The short summary is that our main contributions relative to the previous literature are (i) to introduce the concept of nodal statistics,
(ii) to axiomatize centrality measures
via a common set of axioms showing how they process nodal statistics via exponential weighting,
(iii) to use this to develop a taxonomy of centrality
measures in terms of what information about a nodes' position that they take into account, and on how they weight that information,
(iv) and to provide some further properties that single out some centrality measures and to understand which sorts of trees are such that the
nodal statistics coincide.

\section{Definitions}

\subsection{Background Definitions and Notation}

We consider a network on $n$ nodes indexed by $i\in \{1,2,\ldots n\}$.

A \emph{
network} is a graph, represented by its adjacency matrix $\mathbf{g}\in \Re^{n\times n}$,
where $g_{ij}\neq 0$ indicates the existence of an edge between nodes $i$ and $%
j$ and $g_{ij}=0$ indicates the absence of an edge between the two
nodes.

Our characterization results apply to both directed and undirected versions of networks, and also allow for weighted networks and even signed networks (as the main characterization theorems allow for arbitrary signs and values for links).
Although the results hold without restrictions,  some particular centrality measures are based on unsigned and unweighted networks, and so for the definitions of centrality measures below, we refer to adjacency matrices that are nonnegative and unweighted.\footnote{Some centrality measures proscribe self-loops and so we can adopt the convention that $g_{ii}=0$; but again, the main results do not require such an assumption.}

Let $G(n)$ denote the set of admissible networks on $n$ nodes.

The \emph{degree} of a node $i$ in a undirected network $g$, denoted $d_i(\mathbf{g})= |\{ j:  g_{ij} \neq 0\}|$, is the number of edges
involving node $i$.  (In the case of a directed network, this is outdegree and there is a corresponding indegree defined by  $|\{ j:  g_{ji} \neq 0\}|$.)

A \emph{walk} between $i$ and $j$ is a
succession of (not necessarily distinct) nodes $i=i^0, i^1,...,i^M=j$ such
that $g_{i^mi^{m+1}} \neq 0$ for all $m=0,\ldots ,M-1$.
A \emph{path} in $g$ between two nodes $i$ and $j$  is a
succession of distinct nodes $i=i^0, i^1,...,i^M=j$ such that $%
g_{i^mi^{m+1}}  \neq 0$ for all $m=0,\ldots ,M-1$. Two nodes $i$ and $j$ are
connected (or path-connected) if there exists a path between them.

In the case of an unweighted network, a \emph{geodesic} (shortest path)
from node $i$ to node $j$ is a path such that no other path between them involves a
smaller number of edges.

The \emph{distance} between nodes $i$ and $j$, $%
\rho_\mathbf{g}(i,j)$ is the number of edges involved in a geodesic  between $i$
and $j$, which is defined only for pairs of nodes that have a path between them and may be taken to be $\infty$ otherwise.
The number of geodesics between $i$ and $j$ is denoted $%
\nu_\mathbf{g}(i,j)$.
We let $\nu_\mathbf{g}(k:i,j)$ denote the number of geodesics between $%
i $ and $j$ involving node $k$.

It is useful to note that in the case of  unweighted and unsigned graphs, the elements
of the $\ell$-th power of $\mathbf{g}$, denoted $\mathbf{g}^\ell$, have a straightforward
interpretation: $g^\ell_{ij}$ counts the number of (directed) walks of length $\ell$
from node $i$ to node $j$.

We let $n^{\ell}_i(\mathbf{g})$ denote the number of nodes at distance $\ell$ from $i$ in network $g$:   $ n_i^\ell(\mathbf{g})= |\{j: \rho_\mathbf{g}(i,j)=\ell \}|$.

For the case of undirected, unweighted, unsigned networks,
a \emph{tree} is a graph such that for any two nodes $i,j$ there is a unique
path between $i$ and $j$. A tree can be oriented by selecting one node $i^0$
(the root) and constructing a binary relation $ \succ^d $ as follows: For all nodes such
that $g_{i^0i}=1$, set $i^0 \succ^d i$.   Next, for each pair of nodes $i$ and $j$ that are distinct from $ i^0$,
say that $i \succ^d j$ if $g_{ij}=1$ and the geodesic from $i$ to $i^0$ is
shorter than the geodesic from $j$ to $i^0$.   If $i \succ^d j$, then $i$ is called the \emph{direct
predecessor} of $j$ and $j$ is called a \emph{direct successor} of $i$.  The
transitive closure of the binary relation $\succ^d$ defines a partial order $%
\succ$, where if $i \succ j$ then we say that $i$ is a predecessor of $j$ and $j$ a
successor of $i$, in the oriented tree.

Let $\lambda^{max}(\mathbf{g})$ denote the
largest right-hand-side eigenvalue of a nonnegative $\mathbf{g}$.

\subsection{Some Prominent Centrality Measures}
\

A centrality measure is a function $\mathbf{c}: G(n) \rightarrow \Re^n$, where $%
c_i(\mathbf{g})$ is the centrality of node $i$ in the social network $\mathbf{g}$.\footnote{We define centrality measures as cardinal functions, since that is the way they are all defined in the literature, and are typically used in practice.
Of course, any cardinal measure also induces an ordinal ranking, and sometimes cardinal measures are used to identify rankings.}
Here are some of the key centrality measures from the literature.\footnote{For more background on centrality measures see
\citet[Chapter 4]{wassermanf1994}, \citet{borgatti2005}, \citet[Chapter 2.2]{jackson2008}, and \citet{jackson2020}.}

\bigskip

\noindent \textbf{Degree centrality}

Degree centrality measures the number
of edges of node $i$,
${d_i(\mathbf{g})}$.  We can also normalize by the maximal possible degree, $n-1$, to obtain a number between $0$ and $1$:
$
c^{deg}_i(\mathbf{g}) = \frac{d_i(\mathbf{g})}{n-1}.
$
Degree centrality is an obvious centrality measure, and gives some
insight into the connectivity or `popularity' of node $i$, but misses potentially
important aspects of the architecture of the network and a node's position in the network.\footnote{In the case of directed networks, there are both indegree and outdegree versions, which have different interpretations as to how much node $i$ can either receive or broadcast, depending on the direction.}

\bigskip

\noindent \textbf{Closeness centrality}

Closeness centrality is based on the
network distance between a node and each other node.  It extends
degree centrality by looking at neighborhoods of all radii.  The input into measures of closeness centrality
is
the list of distances between node $i$ and other nodes $j$ in the
network, $\rho_\mathbf{g}(i,j)$. There are different variations of closeness centrality
based on different functional forms. The measure proposed by \citet{bavelas1950} and \citet{sabidussi1966},
is based on distances between node $i$ and all other nodes, $\sum_j
\rho_\mathbf{g}(i,j)$.  In that measure a higher score indicates a lower
centrality.    To deal with this inversion, and also to deal with the fact that this distance becomes infinite if nodes belong to two
different components,  \citet{sabidussi1966} proposed a centrality measure of $\frac{1}{\sum_j \rho_\mathbf{g}(i,j)}$.
One can also normalize that measure so that the highest possible
centrality measure is equal to $1$, to obtain the closeness
centrality measure,
\[
c^{cls}_i(\mathbf{g}) = \frac{n-1}{\sum_{j\neq i} \rho_\mathbf{g}(i,j)}.
\]

An
alternative measure of closeness centrality,  harmonic centrality (e.g., see \cite{rochat2009,garg2009}),
aggregates distances differently.  It aggregates the sum of
all inverses of distances, $\sum_j \frac{1}{\rho_\mathbf{g}(i,j)}$.
This avoids having a few nodes for which there is a large or infinite distance drive the measurement.
This measure can also be
normalized so that it spans from 0 and 1, and one obtains
\[
c^{cl}_i(\mathbf{g}) = \frac{\sum_\ell \frac{1}{\ell} | \{ j : \rho_\mathbf{g}(i,j)=\ell\} | }{n-1} = \frac{1%
}{n-1} \sum_{j \neq i} \frac{1}{\rho_\mathbf{g}(i,j)}.
\]

\bigskip

\noindent \textbf{Decay centrality}

Decay centrality proposed by \citet{jackson2008} is a measure of
distance that takes into account the decay in traveling along shortest paths
in the network. It reflects the fact that information traveling along
paths in the network
may be transmitted stochastically, or that other values or effects transmitted along paths in the network
may decay, according to a
parameter $\delta$.   Decay centrality is defined as
\[
c^{\delta}_i(\mathbf{g}) = \sum_{\ell \leq n-1}\delta^\ell
n^{\ell}_i(\mathbf{g}) .
\]

As $\delta$ goes to $1$, decay centrality measures
the size of the component in which node $i$ lies.  As $\delta$ goes to $0$,
decay centrality becomes proportional to degree centrality.\footnote{Decay centrality is also defined for
$\delta \notin [0,1]$, but then the interpretation of it as capturing decay is no longer valid.}

\bigskip

\noindent \textbf{Katz-Bonacich centrality}

\citet{katz1953} and \citet{bonacich1972,bonacich1987}
proposed a measure of prestige or centrality based on the number of
walks emanating from a node $i$. Because the length of walks in a graph is
unbounded, Katz-Bonacich centrality requires a discount factor
-- a factor $\delta$ between $0$ and $1$ -- to compute the discounted sum of walks
emanating from the node. Walks of shorter length are evaluated at an exponentially higher
value than walks of longer length.\footnote{In the limit, as $\delta\rightarrow 0$, this places weight only on shortest paths,
and then becomes closer to decay centrality, at least in trees. }
In particular, the centrality score for node $i$ is based on
counting the total number of walks from it to other nodes, each exponentially
discounted based on their length:
\[c^{KB}_i(\mathbf{g}, \delta) = \sum_\ell
\delta^\ell \sum_j g^\ell_{ij}.
\]
In matrix terms (when $\mathbf{I} - \delta \mathbf{g}$ inverts):\footnote{ $\mathbf{1}$ denotes the
$n$-dimensional vector of $1$s, and $\mathbf{I}$ is the identity matrix.  Invertibility holds for small enough $\delta$ (less than the inverse of the magnitude of the largest
eigenvalue).}$^,$\footnote{ In a variation proposed by Bonacich there is a second parameter $\beta$ that rescales:
$
\mathbf{c}^{KB} (\mathbf{g}, \delta, \eta) = (\mathbf{I} - \delta \mathbf{g})^{-1} \beta \mathbf{g}
\mathbf{1}.
$  Since the scaling is inconsequential, we ignore it.}
$
\mathbf{c}^{KB }(\mathbf{g},\delta) = \sum_{\ell=1}^{\infty} \delta^\ell
\mathbf{g}^\ell \mathbf{1} = (\mathbf{I} - \delta \mathbf{g})^{-1} \delta \mathbf{g%
} \mathbf{1}.
$

\bigskip

\noindent \textbf{Eigenvector centrality}

Eigenvector centrality, proposed
by \citet{bonacich1972}, is a related measure of prestige. It relies on
the idea that the prestige of node $i$ is related to the prestige of her
neighbors.  Eigenvector centrality is computed by assuming
that the centrality of node $i$ is proportional to the sum of centrality of
node $i$'s neighbors: $\lambda c_i = \sum_j g_{ij} c_j$, where $\lambda$ is
a positive proportionality factor. In matrix terms, $\lambda \mathbf{c} =
\mathbf{g} \mathbf{c}$. The vector $c^{eig}_i(\mathbf{g})$ is thus the
right-hand-side eigenvector of  $\mathbf{g}$ associated with the
eigenvalue $\lambda^{\max}(\mathbf{g})$.\footnote{ $\lambda^{\max}(\mathbf{g})$ is positive when $\mathbf{g}$ is nonzero (recalling that it is a nonnegative matrix), the associated vector is nonnegative, and for a connected network the associated
eigenvector is positive and unique up to a rescaling (e.g., by the Perron-Frobenius Theorem). }

The eigenvector centrality of a node is thus self-referential, but has a well-defined fixed point. This notion of centrality is
closely related to ways in which scientific journals are ranked based on citations, and also relates
to influence in social learning.

\bigskip

\noindent \textbf{Diffusion centrality}

Diffusion centrality, proposed by
\citet*{banerjeecdj2013},\footnote{This is related in spirit to basic epidemiological models (e.g, see \cite{bailey1975}),
as well as the cascade model of \citet*{kempekt2003} that allowed for thresholds of adoption (so that an agent cares
about how many neighbors have adopted).   A variation of the cascade model leads to a centrality measure introduced by \cite*{limot2015} called
cascade centrality,
which is related to the communication centrality of \citet{banerjeecdj2013} and the decay centrality of \citet{jackson2008}.
Diffusion centrality differs from these other measures in that it is based on walks rather than paths, which makes it easier to relate to Katz-Bonacich centrality and eigenvector centrality as discussed in \citet{banerjeecdj2013} and formally shown in \citet*{banerjeecdj2019}.
Nonetheless, diffusion centrality is representative of a  class of measures built on the premise of
how much diffusion one gets from various nodes, with variations in how the process is modeled (e.g., see \cite{bramoulleg2018}).
These are also used as inputs into other measures, such as that of \citet{kermani2015}, which combine information from a variety of centrality measures.}
is based on a dynamic diffusion process starting at node $i$.  In period $1$,
$i$ passes a piece of information to each neighbor with a probability $\delta$.
 In any arbitrary period $\ell$, nodes that received any information at time $\ell-1$ pass each
 piece of information that they have received onwards with probability $\delta$ to each of their neighbors. At period $L$,
the expected number of times that agents have been contacted is computed
using the number of walks
\[
c^{dif}_i(\mathbf{g}, \delta,L) = \sum_{\ell=1}^L  \sum_j
\delta^\ell g^\ell_{ij}.
\]
In matrix terms,
$
\mathbf{c}^{dif} (\mathbf{g},\delta,L) = \sum_{\ell=1}^L \delta^\ell \mathbf{g}%
^\ell \mathbf{1}.
$

If $L=1$, diffusion centrality is proportional to degree
centrality. As $L \rightarrow \infty$, $c^{dif}_i$ converges to
Katz-Bonacich centrality  whenever $\delta$ is smaller than
the inverse of the largest eigenvalue, $1/\lambda^{\max}(\mathbf{g})$.  \citet*{banerjeecdj2013,banerjeecdj2019} show that
diffusion centrality converges to eigenvector
centrality as $L$ grows whenever $\delta$ is larger than
the inverse of the largest eigenvalue, $1/\lambda^{\max}(\mathbf{g})$.

\bigskip

\noindent \textbf{Betweenness centrality}

Freeman's \citeyearpar{freeman1977} betweenness centrality measures
the importance of a node in connecting other nodes in the network. It
considers all geodesics between two nodes $j,k$ different from $i$
which pass through $i$. Betweenness centrality thus captures the role of an
agent as an intermediary in the transmission of information or resources
between other agents in the network. As there may be multiple geodesics connecting $j$ and $k$, we
need to keep track of the fraction of geodesic paths passing through $i$, $%
\frac{\nu_\mathbf{g}(i:j,k)}{\nu_\mathbf{g}(j,k)}$. The betweenness centrality measure proposed by
\citet{freeman1977} is
\[
c^{bet}_i(\mathbf{g}) = \frac{2}{(n-1)(n-2)} \sum_{(j,k), j\neq i,k
\neq i} \frac{\nu_\mathbf{g}(i:j,k)}{\nu_\mathbf{g}(j,k)}.
\]

Betweenness centrality weights all geodesics equally, regardless of how far away the pair of nodes $j,k$ are away from each other, or how many other ways of reaching
each other that they have.
A version that only counts pairs that are directly connected to the given node $i$, was proposed by \cite{jackson2020} under the name of
the {\bf Godfather Index} or Godfather centrality.  The idea is that $i$ can connect pairs of $i$'s friends that are not directly connected to each other.
It takes the simple form:
\[
GF_i (\mathbf{g})= \sum_{k>j}  g_{ik} g_{ij} ( 1- g_{kj})  =   |\{k\neq j: g_{ik}=g_{ij}=1, g_{kj}=0\}|.
\]
It has an inverse relationship to clustering, but weighted by the number of pairs of a node's neighbors.
\[
GF_i(\mathbf{g}) =  (1-clust_i(\mathbf{g})) d_i(\mathbf{g})(d_i(\mathbf{g})-1)/2,
\]
where $clust_i(\mathbf{g})$ is the clustering of node $i$  --  the fraction of pairs of $i$'s neighbors who are connected to each other:
$\sum_{kj\in N_i(\mathbf{g}), k<j } \frac{g_{kj}}{d_i(\mathbf{g})(d_i(\mathbf{g})-1)/2}$.

There are other variations that one can consider.
For instance, instead of only counting immediate neighbors, one can instead weight pairs of other nodes by their distance:\footnote{See \cite{ercsey2012} for some truncated measures.}
$
\sum_{(j,k): j\neq i \neq k \neq j
}  \delta^{\ell_{\mathbf{g}}(j,k)}\frac{\nu_{\mathbf{g}}(i:j,k)}{\nu_{\mathbf{g}}(j,k)},
$
for some $\delta\in [0,1]$.

For example, in a setting where intermediaries
connect buyers and sellers in a network, the number of intermediaries on a
geodesic matters, as intermediaries must share surplus along the path.
In that case, it is useful to consider a variation on betweenness centrality where the
length of the geodesic paths between any two nodes $j$ and $k$ is taken into
account.

Given the number of centrality measures, we do not define them all, but there are many other variations on the above definitions, such as PageRank which is related to Katz-Bonacich and Eigenvector centralities.

\section{Developing a Taxonomy of Centrality Measures}

In this section, we introduce the notion of `nodal statistics' - vectors of data capturing a facet
of the position of a node in the network -- as well as an aggregator that transforms
those vectors of data into centrality measures.   We then characterize how
centrality measures can be seen as processing weighted nodal statistics.   We first introduce nodal
statistics, and then discuss the characterization to show that most centrality measures can be seen
as additively separable weighted sums of nodal statistics.

\subsection{Nodal Statistics}

A \emph{nodal statistic}, $s_i(\mathbf{g})$, is a vector of data describing the position of node $i$ in the network $\mathbf{g}$.
These lie in some Euclidean space, $\Re^L$, where $L$ may be finite or infinite.
We take the vector of all 0's (usually an isolated node, or an empty network) to be a feasible statistic.

Because
networks are complex objects, nodal statistics are useful, as they allow an
analyst to reduce the complexity of a network into a (small) vector of
data. Different nodal statistics capture different aspects of a node's position in
a network.

Standard centrality measures use nodal statistics that pay attention only to the network and not
on the identity of the nodes, as captured in the following property.

For a permutation $\pi$ of $\{1,\ldots ,n\}$,
let $\mathbf{g}\circ\pi$ be defined by $(\mathbf{g}\circ\pi )_{ij} = {g}_{\pi(i)\pi(j)}$

A nodal statistic is {\sl anonymous} if for any permutation $\pi$ of $\{1,\ldots ,n\}$,
$s_i(\mathbf{g}) = s_{\pi(i)} ( \mathbf{g}\circ\pi)$.

\subsubsection{Some Prominent Nodal Statistics}

Several nodal statistics are fundamental.

\bigskip

\noindent \textbf{The neighborhood statistic}, $n_i(\mathbf{g})=
(n_i^1(\mathbf{g}),\ldots,n_i^\ell(\mathbf{g}),\ldots,n_i^{n-1}(\mathbf{g}))$, is a {\sl path-based} vector counting the number of nodes at distance $%
\ell=1,2,\ldots,n-1$ from a given node $i$.\footnote{This concept is first defined in \cite{nieminen1973} in discussing
a directed centrality notion, and who refers to the neighborhood statistic as the subordinate vector.}

The neighborhood statistic measures how quickly (in terms of path length) node $i$ can reach
the other nodes in the network.

\bigskip

\noindent \textbf{The degree statistic}, $d_i(\mathbf{g})=
n^1_i(\mathbf{g})$, counts the connections of a given node $i$.

This is a truncated version of the neighborhood statistic.

\bigskip

\noindent \textbf{The walk statistic}, $%
w_i(\mathbf{g})=(w_i^1(\mathbf{g}),\ldots,w_i^\ell(\mathbf{g}),\ldots)$, is an infinite vector counting the
number of walks of length $\ell=1,2,\ldots$ emanating from a given node $i$. Using the
connection between number of walks and iterates of the adjacency matrix, $%
w_i^\ell (\mathbf{g}) = \sum_j (g^\ell)_{i,j}$.

The main difference from the neighborhood statistic is that it
keeps track of multiplicities of routes between nodes and not just shortest paths, and thus is useful in capturing
processes that may involve random transmission.

\bigskip

\noindent \textbf{The intermediary statistic},
$I
_{i}(\mathbf{g})=(I_{i}^{1}(\mathbf{g}),\ldots,I_{i}^{\ell}(\mathbf{g}),\ldots, I_{i}^{n-1}(\mathbf{g}))$, is a vector counting the normalized
number of geodesics of length $\ell=1,2,\ldots $ which contain node $i$. For
any pair $j,k$ of nodes different from $i$, the normalized number of
geodesic paths between $i$ and $j$ containing $i$ is given by the proportion
of geodesics passing through $i$, $\frac{\nu _\mathbf{g}(i:j,k)}{\nu_\mathbf{g} (j,k)}$.
Summing across over all pairs of nodes $j,k$ different from $i$ who are at distance $\ell$ from each other:  $%
I_{i}^{\ell}=\sum_{jk:\rho_\mathbf{g} (j,k)=\ell, j\ne i, k\ne i}\frac{\nu_\mathbf{g}(i:j,k)}{\nu_\mathbf{g} (j,k)}$.

The
intermediary statistic measures how important node $i$ is in connecting other
agents in the network.

\subsubsection{Ordering Nodal Statistics}

It is often useful to compare two vectors of some nodal statistic:   for instance how does the neighborhood statistic for some node $i$ compare to that of some
other node $i'$?

There are various ways to partially order vectors depending on the application, but given that
the applications all involve vectors in some Euclidean space, the default is to use the Euclidean partial order,
so that $s\geq s'$ if it is at least as large in every entry.

Although the $\geq$ partial order suffices for our main characterization theorems,
in some cases it is useful to also compare the vectors of nodal statistics using other partial orders that are finer and make orderings in cases in which the Euclidean ordering does not.
For example, because the total number of nodes in a connected network is fixed, and the neighborhood
statistic measures the \emph{distribution} of nodes at different distances
in the network, the statistics of two different nodes in a network according to the neighborhood statistic will not be comparable via the Euclidean partial ordering unless they are equal.

Thus, a natural partial order to compare vectors associated with the neighborhood statistic is based on \emph{%
first order stochastic dominance}.\footnote{The entries of the $s$'s may not sum to one, so this is not always a form of {\sl stochastic} dominance, but it is
defined analogously.}
We say that $s_i \succeq
s_i^{\prime }$ if for all $t$, $\sum_{\ell=1}^t s_i^\ell \geq
\sum_{\ell=1}^t s_i^{\prime \ell}$.
This induces a strict version: $s_i \succ
s_i^{\prime }$ if $s_i \succeq s_i^{\prime }$ and $%
s_i^\prime \not\succeq s_i$.
In other words, a statistic $%
s_i$ dominates  $s_i^{\prime }$ if, for
any distance $t$, the number of nodes at distance less than $t$ under $%
s_i$ is at least the number of nodes at distance less than $t$ in
$s_i^{\prime }$.


\subsection{A Characterization of Many Centrality Measures}

We now characterize all of the centrality measures that we have discussed above - i.e., the canonical measures from the literature.

We first present two elementary axioms
that make clear that the centrality of a node $i$ depends only on a node's position and the information
contained in some nodal statistic $s_{i}$.

\begin{axiom}[Anonymity]
\label{symmetry} A centrality measure $c$ is anonymous if for any bijection $\pi $ on $\{1,\ldots ,n\}$ and all $i$: $%
c_{i}\left( \mathbf{g}\right) =c_{\pi (i)}( \mathbf{g} \circ\pi)$.
\end{axiom}

Anonymity guarantees that a centrality measure does not depend on the
identity of a node, but instead on its position in the network.  The next axiom
states that the information about position in the network is captured via a nodal statistic, where higher statistics
correspond to higher centrality.

\begin{axiom}[Monotonicity]
\label{monotonicity} A centrality measure $c$ is monotonic relative to a nodal statistic $s_{i}$ if $s_{i}(\mathbf{g})\geq
s_{i}(\mathbf{g}^{\prime })\ $ implies that $c_{i}\left( \mathbf{g}\right) \geq c_{i}(\mathbf{g}^{\prime })$.
\end{axiom}

Monotonicity connects a centrality measure with a nodal statistic: a
node $i$ is at least as central in social network $\mathbf{g}$ than in social network $%
\mathbf{g}^{\prime }$ if the nodal statistic of $i$ at $g$ is at least as large as the nodal
statistic of $i$ at $\mathbf{g}^{\prime }$.


Under monotonicity and anonymity, the centrality of node $%
i $ only depends on a statistic $s_{i}$, as shown by the following lemma.

Let us say that a function $\mathcal{C}:\Re^L\rightarrow \Re$ is {\sl monotone} if $\mathcal{C}(s)\geq \mathcal{C}(s')$ whenever $s \geq s'$.

We say that a centrality measure is {\sl representable} relative to a nodal statistic $s$ of dimension $L$ if there exists a function  $\mathcal{C}:\Re^L\rightarrow \Re$,  for which $c_{i}(\mathbf{g})=\mathcal{C%
}(s_{i}(\mathbf{g}))$ for all $i$ and $g$.

\begin{lemma}
\label{psi} A centrality measure $c$ is anonymous and satisfies monotonicity relative to some anonymous nodal statistic $s$  if and only if
$c$ is representable relative to the anonymous $s$ by a monotone $\mathcal{C}$.
\end{lemma}

Note that Lemma \ref{psi} implies that the function $\mathcal{C}$ is independent of the index of node $i$.

In what follows, we normalize centrality measures so that $c_{i}(\mathbf{g})=\mathcal{C%
}(s_{i}(\mathbf{g}))=0$ if $s_i(\mathbf{g})=\mathbf{0}$.
This is without loss of generality, as any centrality measure can be so normalized simply by subtracting off
this value everywhere.

We next provide characterizations of increasingly narrow classes of centrality measures by
imposing increasingly strong axioms on how they aggregate nodal statistics.  The standard measures we have defined above all fall into the narrowest class.

We remark that once we have shown that the centrality problem is one of processing nodal statistics,
then it has some parallels to
defining a utility function over a stream of intertemporal consumptions,
as in the classic studies by \citet{koopmans1960} and \citet{debreu1960} in mathematical economics.
Characterizations from that literature have loose parallels here,
although with some important differences in both details and meaning.

\begin{axiom}[Independence]
\label{independence} A function $\mathcal{C}:\Re^L\rightarrow \Re$ satisfies independence if
\[
\mathcal{C}\left( s_{i}\right) -\mathcal{C}\left( s_{i}^{\prime
}\right) = \mathcal{C}\left( s_{i}^{\prime\prime}\right) -\mathcal{C}\left( s_{i}^{\prime\prime\prime }\right).
\]
for any $s_i$,  $s_i^\prime$, $s_{i}^{\prime\prime }$, and $s_{i}^{\prime\prime\prime }$ - all in $\Re^L$ - for which
there exists some $\ell$ such that $s_{i}^{k\prime\prime}=s_{i}^{k}$ and $s_{i}^{k\prime\prime\prime }=s_{i}^{k\prime}$ for all $k\neq \ell$,
and $s_{i}^{\ell}=s_{i}^{\ell \prime }=a$ while $s_{i}^{\ell\prime\prime}=s_{i}^{\ell\prime\prime\prime }=b$ (so, we equally change entry $\ell$ for both statistics without changing any other entries).
\end{axiom}

Independence
requires that, whenever a component of two nodal statistics are
equal, the difference in centrality across those two nodal statistics does not depend on the level of that
component.  
It implies that a centrality measure is an additively
separable function of the elements of the nodal statistic.

\begin{theorem}
\label{decay1}  A centrality measure $c$ is representable relative to an anonymous nodal statistic $s$ by a monotone $\mathcal{C}$ (Lemma \ref{psi})
that satisfies independence,
if and only if there exist a set of monotone functions $F^{\ell}:\mathbb{%
R}\rightarrow \Re$ with $F^{\ell}(0)=0$, such that%
\begin{equation}
c_{i}\left( g\right) =\mathcal{C}(s_{i}\left( g\right)
)=\sum_{\ell=1}^{L}F^{\ell}(s_{i}^{\ell}\left( g\right) ).  \label{decay_eqn1}
\end{equation}
\end{theorem}

Although Theorem \ref{decay1} shows that independence (together with monotonicity and anonymity) implies that a centrality measure is additively
separable, it provides no information on the specific shape of the functions
$F^{\ell}$.
Additional axioms tie down the functional
forms.

A recursive axiom, in the spirit of an axiom from \citet{koopmans1960}, implies that a centrality measure has an exponential aspect to it.

\begin{axiom}[Recursivity]
\label{recursive} A function $\mathcal{C}:\Re^L\rightarrow \Re$ is recursive if for any $\ell< L$ and $s_{i},s_{i}^{\prime }$ (which are $L$ dimensional\footnote{If $L=\infty$, then
when writing $s_{i}^{\ell+1},\ldots,s_{i}^{L} ,0,\ldots, 0 $ below simply ignore the trailing 0's.} )
\begin{equation*}
\frac{\mathcal{C}\left( s_{i}\right) -\mathcal{C} \left(
s_{i}^{1},\ldots,s_{i}^{\ell} ,0,\ldots, 0\right) }{\mathcal{C}\left(
s_{i}^{\prime }\right) -\mathcal{C}\left( s_{i}^{\prime 1},\ldots,s_{i}^{\prime\ell},0,\ldots, 0\right) }=\frac{\mathcal{C} \left(
 s_{i}^{\ell+1},\ldots,s_{i}^{L},0,\ldots,0\right) }{\mathcal{C}
\left( s_{i}^{\prime\ell+1 },\ldots,s_{i}^{\prime L} ,0,\ldots, 0\right) }.
\end{equation*}
\end{axiom}

Recursivity requires that the calculation being done based on later stages of the nodal statistic look
`similar' (in a ratio sense) to those done earlier in the nodal statistic.
For instance, $\mathcal{C}\left( s_{i}\right) -\mathcal{C} \left(
s_{i}^{1},\ldots,s_{i}^{\ell} ,0,\ldots, 0\right)$ captures what is added by the nodal statistic beyond the $\ell$-th entry, and the requirement is
that it operate similarly to how the first $L- \ell$ entries are treated.   The axiom does not require equality, but just that the relative calculations are similar, and hence the ratio component of the axiom.    The axiom makes more sense in the context of particular nodal statistics that are tiered in their construction (e.g., neighborhood, walks, etc.) and for which applying a similar calculation from different starting points can be rationalized.

Again, many standard centrality measures are recursive.  Recursivity implies that the
functions $F^{\ell}$ have an exponential structure $F^{\ell}=\delta ^{\ell-1}f,$
for some increasing $f$ and  $\delta >0$:

\begin{theorem}
\label{decay2}  A centrality measure $c$ is representable relative to an anonymous nodal statistic $s$ by a monotone $\mathcal{C}:\Re^L\rightarrow \Re$ (Lemma \ref{psi}) that is recursive and satisfies independence, if and only if there exists an increasing function $f:\mathbb{R}%
\rightarrow \mathbb{R}$ with $f(0)=0,$ and $\delta \geq 0
$ such that%
\begin{equation}
c_{i}\left( g\right) =\mathcal{C}(s_{i}\left( g\right)
)=\sum_{\ell=1}^{L}\delta ^{\ell-1}f(s_{i}^{\ell}\left( g\right) ).
\label{decay_eqn2}
\end{equation}
\end{theorem}

Recursivity requires each dimension of the nodal statistic to be processed according to the same monotone function $f$, and the only difference in how they enter the centrality measure is in how they are weighted - which is according to an exponential function.

Next, we show that an additivity axiom provides a complete characterization that encompasses all the standard centrality measures,
and results in our main characterization.

\begin{axiom}[Additivity]
\label{additivity} A function $\mathcal{C}:\Re^L\rightarrow \Re$ is additive if for any $s_{i}$ and $s_{i}^{\prime }$ in $\Re^L$:
\begin{equation*}
\mathcal{C}\left( s_{i}+s_{i}^{\prime }\right) =\mathcal{C}(s_{i})+\mathcal{C%
}(s_{i}^{\prime }).
\end{equation*}
\end{axiom}

Additivity is another axiom that is generally satisfied by centrality measures.
It clearly implies independence,\footnote{It also implies monotonicity, but since we use monotonicity
to establish the aggregator function on which additivity is stated, we maintain it as a separate condition
in the statement of the theorem.} and results in the following characterization.

\begin{theorem}
\label{decay3} A centrality measure $c$ is representable relative to an anonymous nodal statistic $s$ by a monotone $\mathcal{C}:\Re^L\rightarrow \Re$ (Lemma \ref{psi}) that is recursive and additive, if and only if
there exists $\delta \geq 0$ and $a\geq 0$ such that%
\begin{equation}
c_{i}\left( g\right) =\mathcal{C}(s_{i}\left( g\right)
)=a \sum_{\ell=1}^{L}\delta ^{\ell-1}s_{i}^{\ell}\left( g\right).
\label{decay_eqn3}
\end{equation}
\end{theorem}

Theorem \ref{decay3} shows that monotonicity, anonymity, recursivity, and additivity, completely tie down that a centrality measure must be the
discounted sum of successive elements of some nodal statistic. Conversely, all centrality measures which can be expressed as discounted sums
of some nodal statistic can be characterized by the axioms of
monotonicity, anonymity, recursivity and additivity.

It is important to emphasize that additivity and recursivity tie down the manner in which nodal statistics are processed to be via an additively separable and linear form.  We are not judging these axioms as being either appealing or unappealing, but instead pointing out that
they have been implicitly assumed when people have defined each of the standard centrality measures.

\subsection{A Taxonomy of Centrality Measures}

Theorem \ref{decay3}
shows that when examining prominent centrality measures -- which all satisfy the axioms of monotonicity, anonymity, recursivity and additivity --
we distinguish them by which nodal statistic they use, and how they weight statistics of various distances.
Thus, a main implication of our results above, including Theorem \ref{decay3},
is that a taxonomy of  centrality measures can be built by which of the basic nodal statistics are used and the weighting scheme.

In particular,
there are three main nodal statistics that are used:  the neighborhood (path) statistic, the walk statistic, and the intermediary (geodesic) statistic;
and then they can be weighted in a variety of ways.
This is pictured in Table \ref{taxonomy}.

\begin{table}[h!]
\begin{tabular}{cccccl}
                               &                                &                                & \multicolumn{2}{c}{Weighting}                                                               &                                  \\ \cline{3-6}
\multicolumn{1}{c}{}          &     \multicolumn{1}{c|}{ }                           & \multicolumn{1}{c|}{ }  &  \multicolumn{1}{c|}{ }  & \multicolumn{2}{c|}{Infinite }              \\
\multicolumn{1}{c}{}          &                                & \multicolumn{1}{|c|}{Immediate} & \multicolumn{1}{c|}{Extended}    & \multicolumn{1}{c}{ low $\delta$} & \multicolumn{1}{c|}{ high $\delta$ }             \\ \hhline{ ~ - |=|=|= =|}
\multicolumn{1}{c|}{}     & \multicolumn{1}{c||}{Neighborhood/}     & \multicolumn{1}{c|}{ } & \multicolumn{1}{c|}{{\bf Decay},}  &\multicolumn{2}{c|}{Not }      \\
\multicolumn{1}{c|}{}     & \multicolumn{1}{c||}{Paths}     & \multicolumn{1}{c|}{{\bf Degree}}    & \multicolumn{1}{c|}{{\bf Closeness}}       & \multicolumn{2}{c|}{Applicable}       \\ \cline{2-6}
\multicolumn{1}{c|}{Nodal}   & \multicolumn{1}{c||}{ }     & \multicolumn{1}{c|}{ } & \multicolumn{1}{c|}{ }  &\multicolumn{1}{c|}{ }      &\multicolumn{1}{c|}{{\bf  Eigenvector},  } \\
\multicolumn{1}{c|}{Statistic} & \multicolumn{1}{c||}{Walks}     & \multicolumn{1}{c|}{{\bf Degree}}    & \multicolumn{1}{c|}{{\bf Diffusion}}   & \multicolumn{1}{c|}{  {\bf Bonacich}  }           & \multicolumn{1}{c|}{{\bf Page Rank}} \\ \cline{2-6}
\multicolumn{1}{c|}{}          & \multicolumn{1}{c||}{Intermediary/} & \multicolumn{1}{c|}{ } &  \multicolumn{1}{c|}{ }  &\multicolumn{2}{c|}{Not }      \\
\multicolumn{1}{c|}{}          & \multicolumn{1}{c||}{Geodesics} & \multicolumn{1}{c|}{{\bf Godfather}} & \multicolumn{1}{c|}{{\bf Betweenness}} & \multicolumn{2}{c|}{Applicable}        \\ \cline{2-6}
\end{tabular}
\caption{\label{taxonomy}A Taxonomy of Centrality Measures}
\end{table}

The taxonomy from Table \ref{taxonomy}
helps us understand which centrality measure is appropriate under which circumstances.

Generally, centrality is used as some measure of how influential a node might be in some process, or how well it is connected (e.g., its social capital.
Which measure captures this depends on the process in question.
The nodal statistic and weighting should be matched with that process.

The neighborhood or path statistic is relevant when one the influence of a node depends on its ``reach''.  For instance, in a simple contagion or diffusion process, paths capture
how many other nodes can be reached in how many steps from a given node -- for instance,  as a disease, idea, or meme moves through the network.
The weighting in that case adjusts for how likely or easily the process transitions from one node to the next.   If it is only immediate connections that are important,  then it is the degree statistic and centrality that matter.   If the process can move further, then decay centrality and its truncated variations are appropriate.

The walk statistic is related to the neighborhood statistic, but instead allows the process to travel in cycles.  In social learning settings, in which information can be repeated and ``echo'', and in which
repeatedly hearing things matters,  then walk statistics become more appropriate.   The weighting then captures the probability of transmission.
Walk statistics can also be appropriate in game theoretic settings (e.g., see the survey in \cite{jacksonz2014}), in which people influence each other's behaviors and influence then transmits indirectly.
There, the weighting captures the amount of influence a person has on their neighbors.

The intermediary statistic has a very different interpretation than the other two.  It is appropriate when measuring how important an individual is as a connector between others.
Here, weighting can capture the fact that connecting people directly is of more value than connecting people at a distance.

We remark that eigenvector centrality is actually a limiting case.
In particular, note that if the discount factor $\delta$ is larger than the inverse of the
largest eigenvalue of the adjacency matrix, then there exists $\overline{L}$ large enough such that the
ranking generated by diffusion centrality will remain the same for any $L
\geq \overline{L}$.
In fact, for large enough $L$,  diffusion
centrality converges to  eigenvector
centrality if the discount factor $\delta$ is larger than the inverse of the
largest eigenvalue of the adjacency matrix (see \citet{banerjeecdj2013, banerjeecdj2019}).\footnote{For smaller $\delta$ diffusion centrality coincides with Katz-Bonacich centrality, and
so exactly at the inverse of the largest eigenvalue, Katz-Bonacich and eigenvector centrality converge.  This presumes that there is a unique first eigenvector, which holds if the adjacency matrix is primitive (e.g., see \cite{jackson2008}), which is true as long as there is (directed) path between any pair of nodes -- e.g., if there is a single component in an undirected network.
\cite{bonacich2007} discusses some interesting properties of eigenvector centrality and how it can differ on signed and other networks, which violate these conditions. }
Thus, eigenvector centrality is simply the limit of a centrality measure that satisfies the axioms and has a representation of the form in Theorem \ref{decay3}.

\medskip

We also remark that closeness centrality is the one centrality measure that uses a different weighting scheme.
It still uses the neighborhood statistic, but instead of weighting nodes at different distances exponentially, doing it hyperbolically.   That is, instead of
weighting the neighborhood statistic by $\delta^\ell$, it is weighted by $1/\ell$.   It is then covered under Theorem \ref{decay1} instead of Theorem \ref{decay3}, but can still be viewed as part of the first row -- together with decay centrality --  just with a different weighting scheme.\footnote{Alternatively, we could define a closeness statistic, $cl_i(\mathbf{g})=
(cl_i^1(\mathbf{g}),\ldots, cl_i^\ell(\mathbf{g}),\ldots,cl_i^{n-1}(\mathbf{g}))$, is the vector such that
$
cl_i^\ell(\mathbf{g}) = \frac{n_i^\ell(\mathbf{g}) }{\ell}
$
for each $%
\ell=1,2,\ldots,n-1$, tracking nodes at different distances from a given node $i$, weighted by the inverse of those distances.
and add another row.  But this would build some of the weighting into the nodal statistics, which is cleaner to separate, pedagogically.}

\subsection{Path-Based Nodal Statistics}

Given the extensive use of degree centrality (and variations of closeness and decay), we identify some conditions that tie down the neighborhood statistic.
Not only does this help understand it, but the cycle-independence condition helps us to understand how it differs from the walk statistic.

Given a network $g$, we say that a subnetwork $g'$ is a {\sl minimal $i$-centered subtree} if it is a tree and the path distance between $i$ and $j$ in $g'$ is the same as
it is in $g$ for all $j\neq i$.

Every network $g$ and node $i$ have at least one associated minimal $i$-centered subtree, which is necessarily also an $i$-centered subtree of the component
of $g$ containing $i$.

To understand the idea of $i$-centering, we note that it preserves the distance structure of the network.
This is not true of just any minimal spanning tree.
For instance,
consider a network connecting $n$ agents that consists of a single giant cycle -- essentially a circle.  There are $n$ minimal spanning trees for this network, each found by eliminating a different link.  But depending on which tree one examines, node $i$ could become closer or further from some nodes.  For instance, if we delete the link from $i$ to $i+1$, then even though those two nodes are
actually next to each other in the original network, $i+1$ now appears $n-1$ links away from $i$ in the minimal spanning tree.   If we wish to calculate decay centrality, then we need to pick a spanning tree in which $i$ is in the ``middle'' - so eliminating a link as far as possible from node $i$.  This is accomplished by an $i$-centered tree.

\begin{axiom}[Cycle Independence]
\label{tree1} A centrality measure $c$ is cycle independent if
$c_i(g)=c_i(g')$ for every network $g$, every $i$, and any minimal $i$-centered subtree $g'$.
\end{axiom}

Cycle independence states that centrality can be found by looking just at the tree structure of a network and ignoring additional cycles that exist.

We say that a tree $g$ {\sl dominates} another tree $g'$ for $i$ if the number of nodes within distance $\ell$ of $i$ is weakly greater under $g$ than $g'$ for every
$\ell$.   We say that $g$ {\sl strictly dominates} another tree $g'$ if $g$ dominates $g'$ and the number of nodes within distance $\ell$ of $i$ is strictly greater under $g$ than $g'$ for some $\ell$.

\begin{axiom}[Distance Sensitivity]
\label{tree2}  Centrality measure $c$ is distance sensitive if $g$ and $g'$ are two trees for which
$g$ dominates $g'$ for $i$, then
$c_i(g)\geq c_i(g')$, with strict inequality if the dominance is strict.
\end{axiom}

Distance sensitivity implies that the key information from a tree is the distances of nodes from the root.

Centrality measures that are based on neighborhood statistics also impose additional restrictions on the way in which distances enter calculations.
In particular, they
are separable in the sense that the marginal contribution of a node at a given distance remains constant relative to the rest of the network, as captured in the following
axiom.

\begin{axiom}[Constant Marginal Values]
\label{ax-tree2}  Centrality measure $c$ has constant marginal values if $g$ and $g'$ are two trees for which node $j$ is not involved
and $j$ is at the same distance from $i$ in both $g+kj$ and $g'+k'j$ for some added links $kj$ to $g$ and $k'j$ to $g'$, then
$c_i(g+kj) -c(g)=  c_i(g'+k'j)- c_i(g')$.
\end{axiom}

We remark that there are many variations on a centrality measure that give the same ordering over nodes as a function of the network that differ cardinally.
Thus, without relying on axioms that are explicitly cardinal in nature, dictating how distance or some other aspect of position translates into centrality,
any characterization will be up to some equivalence class, where the actual cardinal values can vary.
This prompts the next definition that captures statistics that are relative re-weightings of the neighborhood statistic.

We say that $s_{i}\left( g\right)$  is a {\sl weighted neighborhood statistic} if there exists a vector of weights $w\in \Re^L_+$ such that
$s_i(g) _\ell =  w_\ell  n_i^\ell (g)$.
It is $\delta$-monotone if $w_\ell \delta^\ell$ is decreasing in $\ell$.

\begin{theorem}
\label{fulldecay} A centrality measure $c$ is cycle independent, distance sensitive, has constant marginal values, and is representable relative to an anonymous nodal statistic $s$ by a monotone $\mathcal{C}:\Re^L\rightarrow \Re$ (Lemma \ref{psi}) that is recursive and additive\footnote{Without the restriction that $L=n-1$ one
can get additional statistics that repeat entries -- for instance instead of having the neighborhood statistics $(n_i^1(g), n_i^2(g), n_i^3(g), \ldots , n_i^{n-1}(g))$,
one can also get other statistics such as
$(n_i^1(g), n_i^1(g), n_i^2(g), n_i^2(g), n_i^3(g), n_i^3(g),\ldots , n_i^{n-1}(g), n_i^{n-1}(g))$ which duplicates entries.
} if and only if
there exists $ \delta > 0$ and $a> 0$ such that%
\begin{equation*}
c_{i}\left( g\right) =\mathcal{C}(s_{i}\left( g\right)
)=a \sum_{\ell=1}^{L}\delta ^{\ell-1}s_{i}^{\ell}\left( g\right),
\end{equation*}
and such that $s_{i}\left( g\right)$ is a weighted neighborhood statistic that is $\delta$-monotone.
\end{theorem}

Theorem \ref{fulldecay} shows that a weighted variation of the neighborhood statistic is at the origin of any centrality measure satisfying (in addition to the recursivity and additivity axioms) cycle independence and distance sensitivity. These two additional axioms guarantee that the centrality measure only depends on information contained in $i$-centered tree subtrees and is sensitive to the distances of paths in these trees - which ties down that the information is equivalent to nodes at various distances, up to some monotone weighting of that information.   When the weights are all one (or all the same), then
the resulting centrality measure is decay centrality.  As the weights vary, then it is a weighted version of decay centrality.
The cycle independence is a key axiom distinguishing decay centrality from other measures in the more general family from Theorem \ref{decay3}.

The next result pushes further, so that all the weight is on direct connections.

\begin{axiom}[Long Distance Insensitivity]
\label{deg1}  Centrality measure $c$ is long distance insensitive if $g$ and $g'$ are two trees for which
$i$ has weakly more links under $g$
than $g'$, then
$c_i(g)\geq c_i(g')$, with strict inequality if $i$ has strictly more links.
\end{axiom}

\begin{theorem}
\label{degree} A centrality measure $c$ is cycle independent, long distance insensitive, has constant marginal values, and is representable relative to an anonymous nodal statistic $s$ by a monotone $\mathcal{C}:\Re^L\rightarrow \Re$ (Lemma \ref{psi}) that is recursive and additive if and only if $c$ is proportional to degree centrality.
\end{theorem}

Theorems \ref{fulldecay} and \ref{degree} offer characterizations of two types of centrality measures.
As with any characterization, one may or may not find the key axioms appealing - which is why they are helpful in crystalizing what is or is not appealing about a measure.
For example, cycle independence may make sense in situations in which shortest paths are important and redundant paths are not useful, but may be less appealing in a setting in which paths
fail with nontrivial probabilities and then cycles have some importance.
We leave it to future research to characterize other nodal statistics, and thus provide a fuller bestiary of axioms.

\subsection{Some Related Literature}

A main contribution of our work is to show that standard centrality
measures can be distinguished along two dimensions:
which nodal statistics are paid attention to when aggregating, and how they are weighted.    

This distinguishes our work from the previous literature.
Let us discuss some of the previous analysis of centrality measures.

This is obviously not the first article or book to note that centrality measures are based on processing some sort information about nodes' positions in a network (e.g., see \cite{jackson2008,schochb2016}).
For instance, \cite{borgattie2006,schochb2016} note that most centrality measures involve working with various aspects of walk counts or path algebras.
Indeed, as is seen above, most of the nodal statistics described above depend in some
way on walks in a network.  \cite{schochb2016} show that centrality measures respect a monotonicity condition relative to the path algebras, so that if one
node dominates another according to a measurement of position, then it will be more central.   We use a related observation as one piece of Lemma \ref{psi}, in order to define
representation of a centrality measure.

The contribution here is that we show not only that standard centrality measures depend on nodal statistics, but also that they all employ a very specific additively
separable exponential weighting method; which then enables our taxonomy.

Centrality measures are also related to other
ranking problems.  For example, ranking
problems have been considered in the contexts of tournaments \citep{laslier1997},
citations across journals \citep{palacios-heurtav2004}, and hyperlinks
between webpages \citep*{page1998}. There is a literature in social choice devoted to
the axiomatization of ranking methods. For example, the Copeland rule (by
which agents are ranked according to their count of wins in a tournament),
-- the equivalent of degree centrality in our setting -- has been
axiomatized by \citet{rubinstein1980}, \citet{henriet1985}, and \citet{brinkg2003}.
\citet{palacios-heurtav2004} axiomatize the invariant solution -- an
eigenvector-based measure on a modified matrix normalized by the number
of citations. Their axiomatization relies on global properties -- anonymity,
invariance with respect to citations intensity. It then introduces an axiom
characterizing the solution for $2 \times 2$ matrices (weak homogeneity) and
a specific definition of reduced games, which together with a consistency
axiom, allows to extend the solution in $2 \times 2 $ games to general
matrices. \citet{slutzkiv2006} propose an alternative axiomatization of
the invariant solution, replacing weak homogeneity and consistency by a weak
additivity axiom. They characterize the invariant solution as the only
solution satisfying weak additivity, uniformity and invariance with respect
to citations intensity. \citet{slutzkiv2006} axiomatize a different
eigenvector centrality measure -- the fair bets solution. The fair bets
solution is the only solution satisfying uniformity, inverse proportionality
to losses and neutrality.   \citet{boldiv2014} axiomatize a variation of closeness centrality that they refer to as
harmonic centrality.

\citet{dequiedtz2014} recently proposed an axiomatization of
prestige network centrality measures, departing from the axioms of \citet{palacios-heurtav2004}
in several directions. As in \citet{palacios-heurtav2004},
their axiomatization relies on the characterization of the
solution in simple situations (in this case stars) and the definition of a
reduced problem such that consistency extends the solution from the simple
situation to the entire class of problems. The reduced game is defined using
the concept of an ``embedded network'': a collection of nodes
partitioned into two groups - one group where a value is already attached to
the node (terminal nodes) and one group where values still have to be
determined (regular nodes). One axiom used is a normalization axiom. Two
axioms are used to determine the solution in the star -- the linearity and
additivity axioms. Consistency is then applied to generate a unique solution
-- the Katz Bonacich centrality measure with an arbitrary parameter $a$.
Replacing linearity and additivity by invariance, \citet{dequiedtz2014}
obtain a different solution in the star network, which extends by
consistency to degree centrality for general situations. Eigenvector
centrality can also be axiomatized using a different set of axioms on the
star network, and adding a converse consistency axiom.

\citet{garg2009}\footnote{Garg's paper was never completed, and so the
axiomatizations are not full characterizations and/or are without proof.  Nonetheless some of the axioms in his paper are of interest.}
proposed different sets of axioms to characterize each of
degree, decay and closeness centralities.  To axiomatize degree centrality, he uses an additivity axiom
across subgraphs -- a much stronger requirement than that discussed here, which makes the
measure independent of the structure of neighborhoods at distance greater
than one. In order to axiomatize decay and degree centrality, Garg
uses an axiom which amounts to assuming that the only relevant
information in the network is the neighborhood statistic. The "breadth first
search" axiom assumes that centrality measures are identical whenever two
graphs generate the same reach statistics for all nodes. A specific axiom of
closeness pins down the functional form of the additively separable
functions so that the closeness centrality measure is obtained. In order to
characterize decay centrality, Garg uses another axiom
which pins down a specific functional form, termed the up-closure axiom.

Finally, \cite{jackson2020} provides some discussion of centrality measures as they relate to social capital, but using a very different taxonomy, and without providing
the characterization here.






\section{When do Centrality Measures Agree?}

In this section, we compare centrality measures both theoretically and via some simulations.
Here, the use of nodal statistics is quite powerful and allows us to precisely characterize when it is that different centrality measures coincide.

\subsection{Comparing Centrality Measures on Trees}

We first focus attention on trees and characterize the class of trees for which all  centrality measures based on the neighborhood statistic coincide, and then characterize the class of trees for which all centrality measures based on the neighborhood, intermediary and walk statistics coincide.

For networks that are {\em not} trees, the characterization of all networks for which centrality measures coincide is an open problem.
The presence of cycles means that nodal statistics that only track neighborhood structures (e.g., just distances) can differ dramatically from statistics that track overall path structure (e.g., numbers of paths connecting nodes).\footnote{\citet*{koenigtz2014}
 prove that degree, closeness, betweenness and eigenvector centrality generate the same ranking on nodes for nested-split graphs, which are a very structured hierarchical form of network (for which all nodal statistics will provide the same orderings, and so the techniques here would provide an alternative proof technique).}

\subsubsection{Monotone Hierarchies}

We define a class of trees that we call \emph{monotone hierarchies}. A tree $\mathbf{g}$ is
a monotone hierarchy if there exists a node $i_0$ (the root) such that the
oriented tree starting at $i_0$ satisfies the following conditions:

\begin{itemize}
\item For any two nodes $i,j$, if the distance between the root and $i$ , $\rho(i)$, is smaller than the distance between the root and $j$, $\rho(j)$, then  $d_i \geq d_j$.

\item For any two nodes $i,j$ such that $\rho(i)=\rho(j)$, if $d_i
> d_j$, then $d_k \geq d_l$ for any successor $k$ of $i$ and any successor $%
l $ of $j$ such that $\rho(k)= \rho(l)$.

\item For any two nodes $i,j$ such that $\rho(i)=\rho(j)$, if $d_i
= d_j$, and $d_k > d_l$ for some successor $k$ of $i$ and any successor $%
l $ of $j$ such that $\rho(k)= \rho(l)$, then $d_k \geq d_l$ for every successor $k$ of $i$ and successor $%
l $ of $j$ such that $\rho(k)= \rho(l)$.

\item All leaves are at the same distance from the root node.\footnote{The other conditions guarantee that all leaves' distances from the root differ by no more than one from each other.  However, a line with an even number of nodes shows that there will be no well-defined root node that is more central than other nodes, and such examples are ruled out by this condition.}
\end{itemize}

In a monotone hierarchy, nodes  further from the root have a
weakly smaller number of successors.
In a monotone hierarchy, different subgraphs may have different numbers of
nodes. However, if at some point, a node $i$ has a larger number of
successors than a node $j$ at the same level of the hierarchy, in the
sub-tree starting from $i$, all nodes must have a (weakly) larger degree
than nodes at the same level of the hierarchy in the sub-tree starting from $%
j$.

\begin{equation*}
\begin{tikzpicture}[inner sep=1mm]
\tikzstyle{place}=[circle,draw=blue!50,fill=blue!20,thick,minimum size=3mm]
\tikzstyle{post}=[<-,shorten >=1pt,>=stealth,semithick] \node at
(8,12)[place](1){0} ; \node at (5,9)[place](2){1} edge [post] node {} (1);
\node at (7,9)[place](3) {2} edge [post] node {}(1) ; \node at (9,9)
[place](4) {2} edge [post] node {}(1) ; \node at (11,9) [place](5) {2} edge
[post] node {}(1) ; \node at (3.5,7) [place](6) {3} edge [post] node {}(2) ;
\node at (4.5,7) [place](7) {4} edge [post] node {}(2) ; \node at (5.5,7)
[place] (8) {4} edge [post] node {}(2) ; \node at (6.5,7) [place] (9) {5}
edge [post] node {}(3) ; \node at (7.5,7) [place] (10) {5} edge [post] node
{}(3) ; \node at (8.5,7) [place] (11) {5} edge [post] node {}(4) ; \node at
(9.5,7) [place] (12) {5} edge [post] node {}(4) ; \node at (10.5,7) [place]
(13) {5} edge [post] node {}(5) ; \node at (11.5,7) [place] (14) {5} edge
[post] node {}(5) ; \node at (2.5,5) [place] (15) {6} edge [post] node {}(6)
; \node at (3.5,5) [place] (16) {6} edge [post] node {}(6) ; \node at
(4.5,5) [place] (17) {7} edge [post] node {}(7) ; \node at (5.5,5) [place]
(18) {7} edge [post] node {}(8) ; \node at (6.5,5) [place] (19) {8} edge
[post] node {}(9) ; \node at (7.5,5) [place] (20) {8} edge [post] node
{}(10) ; \node at (8.5,5) [place] (21) {8} edge [post] node {}(11) ; \node
at (9.5,5) [place] (22) {8} edge [post] node {}(12) ; \node at (10.5,5)
[place] (23) {8} edge [post] node {}(13) ; \node at (11.5,5) [place] (24)
{8} edge [post] node {}(14) ; \node at (2.5,3) [place] (25) {9} edge [post]
node {} (15) ; \node at (3.5,3) [place] (26) {9} edge [post] node {}(16) ;
\node at (4.5,3) [place] (27) {10} edge [post] node {}(17) ; \node at
(5.5,3) [place] (28) {10} edge [post] node {} (18) ; \node at (6.5,3)
[place] (29) {11} edge [post] node{}(19) ; \node at (7.5,3) [place] (30)
{11} edge [post] node{}(20) ; \node at (8.5,3) [place] (31) {11} edge [post]
node{}(21) ; \node at (9.5,3) [place] (32) {11} edge [post] node{}(22) ;
\node at (10.5,3) [place] (33) {11} edge [post] node{}(23) ; \node at
(11.5,3) [place] (34) {11} edge [post] node{}(24) ; \end{tikzpicture}
\end{equation*}


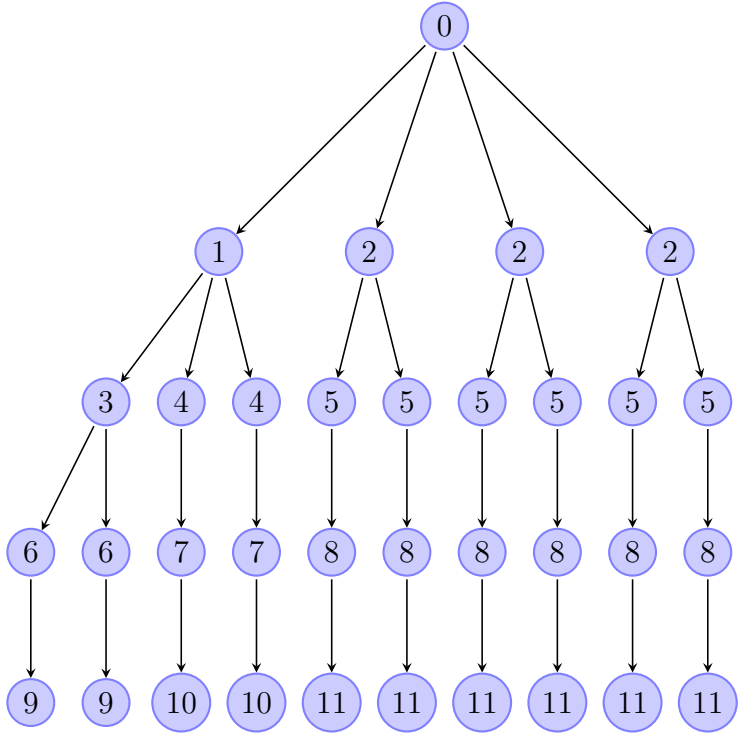
\captionof{figure}{A monotone hierarchy} \label{monhier}

\bigskip

Figure \ref{monhier} displays a monotone hierarchy, with numbers
corresponding to the ranking of nodes. A node has a lower number if it has a higher ranking in the hierarchy. Two nodes have the same number if neither of them is ranked above the other. A node has a higher ranking if it is closer to the root, or, for two nodes at the same level of the hierarchy, if it belongs to a subtree with a larger number of nodes. More generally, we define the following partial order, $i \gtrdot j$, on nodes in a monotone
hierarchy.

\begin{itemize}
\item If $\rho(i) < \rho(j)$, then $i \gtrdot j$.

\item For nodes $i$ and $j$ at the same distance from the root, $\rho(i) = \rho(j)$, we define the condition inductively starting with
nodes at distance one:
\begin{itemize}
\item For $\rho(i) = \rho(j)=1$, if either $d_i > d_j$, or if $d_i=d_j$ and there exist two
successors $k,l$ of $i$ and $j$, respectively, such that $\rho(k) = \rho(l)$ and $d_k
> d_l$, then $i \gtrdot j$.

\item Inductively, in distance from the root, consider $i$ and $j$ such that $\rho(i) = \rho(j)>1$:
\begin{itemize}
\item If there exist two
distinct predecessors $k,l$ of $i,j$, respectively, such that $\rho(k) = \rho(l)$ and $%
k \gtrdot l$, then $i \gtrdot j$.
\item If either $d_i > d_j$, or if $d_i=d_j$ and there exist two
successors $k,l$ of $i$ and $j$, respectively, such that $\rho(k) = \rho(l)$ and $d_k
> d_l$, then $i \gtrdot j$.\footnote{Note that this condition cannot be in conflict with the previous one, as it would violate the ordering of $k$ and $l$.
This latter condition only adds to the definition when $i$ and $j$ have the same immediate predecessor.}
\end{itemize}
\end{itemize}
\end{itemize}

Note that, by the definition of a monotone hierarchy, the only way in which two nodes are not ranked relative
to each other is that they are at the same level and every subtree containing one, and a corresponding subtree containing the other that starts at the same level must
be the homomorphic (the same, ignoring node labels).
In that case we write $i \doteq j$ to represent that neither $i \gtrdot j$ nor $j \gtrdot i$).

Hence, the
ranking $\geqdot$,  $\gtrdot$ with associated $\doteq$, is well defined and gives a complete and transitive ranking of the
nodes.

\bigskip

As we now show, monotone hierarchies are the only trees for which all
centrality measures defined by the neighborhood statistic coincide.

\begin{proposition}
\label{theomonhier} In a monotone hierarchy, for any two nodes $i,j$, $i
\gtrdot j$ if and only if $n_i \succ n_j$ and $i \doteq j$
if and only if $n_i = n_j$. Conversely, if a tree with even diameter and all leaves equidistant from the root\footnote{Without this condition, there are examples of trees that violate being a monotone hierarchy
 because of the leaf condition, but still have all nodes being comparable in terms of their neighborhood structures.} is not a monotone
hierarchy, there exist two nodes $i$ and $j$ such that neither $n_i
\succeq n_j$ nor $n_j \succeq n_i$.
\end{proposition}

\subsubsection{Regular Monotone Hierarchies}

For all centrality measures based on other nodal statistics to coincide too, we need to consider a more
restrictive class of trees, which we refer to as \emph{regular monotone hierarchies}. A
monotone hierarchy is a regular monotone hierarchy if
all nodes at the same distance from the root have the same degree ($d_i=d_j$ if $\rho(i) = \rho(j)$).

Cayley trees are regular monotone hierarchies. Stars and lines
are regular monotone hierarchies. In a regular monotone
hierarchy, all nodes at the same distance from the root are symmetric and
hence have the same centrality. Centrality is highest for the root $i_0$ and
decreases with the levels of the hierarchy. Figure \ref{regmonhier}
illustrates a regular monotone hierarchy.

\begin{equation*}
\begin{tikzpicture}[inner sep=1mm]
\tikzstyle{place}=[circle,draw=blue!50,fill=blue!20,thick,minimum size=3mm]
\tikzstyle{post}=[<-,shorten >=1pt,>=stealth,semithick] \node at
(8,12)[place](1){0} ; \node at (5,9)[place](2){1} edge [post] node {} (1);
\node at (7,9)[place](3) {1} edge [post] node {}(1) ; \node at (9,9)
[place](4) {1} edge [post] node {}(1) ; \node at (11,9) [place](5) {1} edge
[post] node {}(1) ; \node at (4.5,7) [place](6) {2} edge [post] node {}(2) ;
\node at (5.5,7) [place] (7) {2} edge [post] node {}(2) ; \node at (6.5,7)
[place] (8) {2} edge [post] node {}(3) ; \node at (7.5,7) [place] (9) {2}
edge [post] node {}(3) ; \node at (8.5,7) [place] (10) {2} edge [post] node
{}(4) ; \node at (9.5,7) [place] (11) {2} edge [post] node {}(4) ; \node at
(10.5,7) [place] (12) {2} edge [post] node {}(5) ; \node at (11.5,7) [place]
(13) {2} edge [post] node {}(5) ; \node at (4.5,5) [place] (14) {3} edge
[post] node {}(6) ; \node at (5.5,5) [place] (15) {3} edge [post] node {}(7)
; \node at (6.5,5) [place] (16) {3} edge [post] node {}(8) ; \node at
(7.5,5) [place] (17) {3} edge [post] node {}(9) ; \node at (8.5,5) [place]
(18) {3} edge [post] node {}(10) ; \node at (9.5,5) [place] (19) {3} edge
[post] node {}(11) ; \node at (10.5,5) [place] (20) {3} edge [post] node
{}(12) ; \node at (11.5,5) [place] (21) {3} edge [post] node {}(13) ; \node
at (4.5,3) [place] (22) {4} edge [post] node {}(14) ; \node at (5.5,3)
[place] (23) {4} edge [post] node {}(15) ; \node at (6.5,3) [place] (24) {4}
edge [post] node {}(16) ; \node at (7.5,3) [place] (25) {4} edge [post] node
{}(17) ; \node at (8.5,3) [place] (26) {4} edge [post] node {}(18) ; \node
at (9.5,3) [place] (27) {4} edge [post] node {}(19) ; \node at (10.5,3)
[place] (28) {4} edge [post] node {}(20) ; \node at (11.5,3) [place] (29)
{4} edge [post] node {}(21) ; \end{tikzpicture}
\end{equation*}


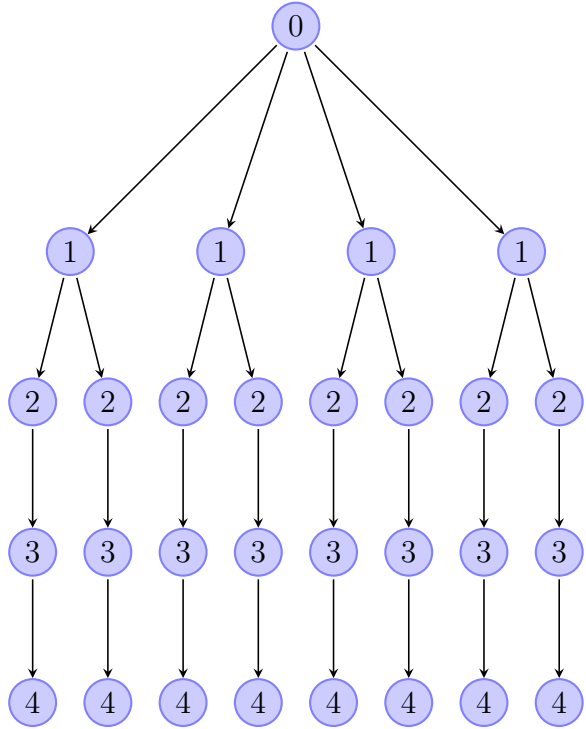
\captionof{figure}{A regular monotone hierarchy} \label{regmonhier}

\bigskip

In this case, the ranking $\gtrdot$ corresponds completely with distance to the root: $i \gtrdot j$ if and only if $\rho(i)
< \rho(j)$ and $i \doteq j$ if and only if $\rho(i)= \rho(j)$.

\begin{proposition}
\label{theoregmonhier2} In a regular monotone hierarchy, $i \gtrdot j$ if and
only if $n_i \succ n_j, I_i \succ I_j$ and $w_i \succ
w_j$; and $i \doteq j$ if and only if $n_i = n_j, I_i =
I_j$ and $w_i = w_j$. For any tree which is not a regular
monotone hierarchy, there exist two nodes $i$ and $j$ and two statistics $s,
s^{\prime } \in \{n,I,w\}$  such that $s_i \succeq s_j$ and $s^{\prime }_j \succ s^{\prime
}_i$.
\end{proposition}

Proposition \ref{theoregmonhier2} shows that, in a regular monotone hierarchy,
{\sl all} centrality measures based on neighborhood, intermediary, and walk statistics
rank nodes in the same order: based on their distance from the root.  This is also true for any other
statistic for which distance from the root orders nodal statistics according to $\succeq$ (and it is hard to think of any natural statistic that would not
do this in such a network).
Conversely, if
the social network is a tree which is not a regular monotone hierarchy,
then the
centrality measures will not coincide. The intuition underlying Proposition \ref%
{theoregmonhier2} is as follows.  In
a regular monotone hierarchy, agents who are more distant from the root have
longer distances to travel to other nodes, are less likely to lie on paths between other
nodes, and have a smaller number of walks emanating from them. Next consider
the leaves of the tree. By definition, they do not sit on any path
connecting other agents and have a intermediary statistic equal to $%
I=(0,0,...0)$. Hence, if centrality measures are to coincide, all leaves
of the tree must have the same neighborhood statistic, a condition which can only
be satisfied in a regular monotone hierarchy. This last argument shows that
centrality rankings based on the intermediary and neighborhood statistic can only
coincide in regular monotone hierarchies.

\subsection{Simulations: Differences in Centrality Measures by Network Type}

Given how extreme the network structures have to be before centrality measures agree,
we can also explore some, how they disagree as a function of the network structure.
We do this in the appendix by simulating networks where we can control the network characteristics,
such as density, homophily, and bridge structures.
Given all of these dimensions, we end up with many different networks on which to compare centrality measures, and so many of the results appear in the appendix.

\section{Concluding Remarks: Potential for New Measures}

Given that our results show that all standard centrality measures are based on the same method of aggregation, and have a parallel to the axioms that characterize time-discounted additively separable utility functions,  there seems to be room
for the development of new measures.  We close with thoughts on such classes of measures.

First, we could allow for more general discounting , with different exponential parameters for different path lengths, in the spirit of hyperbolic discounting. We could also truncate closeness and decay centrality to have a maximal path length $L$, as in diffusion centrality, beyond which effects disappear.

Another class of measures that may be worth exploring in greater detail are those based on power indices from cooperative game theory, with the Shapley value being a prime example.   \cite{myerson1977} adapted the Shapley value to allow for communication structures, \cite{jacksonw1996} adapted the Myerson value to more general network settings, and
\cite{brinkg2000} adapted it for power relationships represented by hierarchies and directed networks.
The Myerson value defined in \cite{jacksonw1996} provides a whole family of centrality measures, as once one ties down how value is generated by the network, it then indicates how much of that value is allocated - or `due' - to each node.  Some
variations of these measures have popped up in the later literature \citet*{gomez2003,michalak2013,molinero2013,ciardiello2018}.
These can be difficult to compute, and in some cases still satisfy variations on the additivity axiom.  It appears that whether or not the additivity axiom would be violated depends on the choice of the value function.\footnote{Even though the Shapley value satisfies an additivity axiom, it is an additivity across value functions and not across nodal statistics; and so does not translate here.}
The choice of value function would be tied to the application.

A different idea, which is a more direct variation on standard centrality measures, is to look at the probability of infecting a whole population starting from some node under some diffusion/contagion process, rather than the expected number of infected nodes, as embodied in the notion of contagion centrality defined by \citet*{limot2015}.   One could also examine whether some given fraction of a population is reached, or whether a nontrivial diffusion is initiated from some node. Additionally, a threshold model where a fraction of neighbors (or an absolute number of neighbors) must be infected for diffusion to happen is also an interesting thing to explore.\footnote{In the case of a threshold model, as multiple seeds are needed to initiate any cascade in many networks, one could construct a centrality measure by assuming that $k$ other seeds are distributed at random on all other nodes, and then examine the marginal value of a particular node.} Although such measures build on the same sorts of models as diffusion and related centralities, they clearly violate the additivity axiom, and so would move outside of the standard classes.   The differences that they exhibit compared to standard measures would be interesting to explore.

Another new class of measures that may be worth exploring involve a multiplicative formulation instead of an additive one.  This would reflect strong complementarities among different elements of the nodal statistics, for instance nodes at various distances.
Given scalars $\alpha_\ell$ and $\beta_\ell$ that capture the relative importance of the different dimensions of the nodal statistics, for instance the role of nodes at various distances from the node in question, we define a new family of centrality measures as follows:
\begin{equation}
c_{i}\left( g\right) =\mathcal{C}(s_{i}\left( g\right)
)=\times_{\ell=1}^{L} (\alpha_{\ell }+ s_{i}^{\ell}\left( g\right))^{\beta_{\ell}}.
\label{new}
\end{equation}
These are a form of  multiplicative measures that parallel the form of some production functions and would capture the idea, for instance, that nodes at various distances are complementary inputs into a production process for a given node.\footnote{It would generally make sense to have the $\beta_\ell$ be a non-increasing function of $\ell$.   The presence of the $\alpha_{\ell}$s ensures that there is no excessive penalty for having $s_i^\ell =0$ for some $\ell$.}
This class of measures could produce different rankings of nodes compared to standard centrality measures, and would capture ideas such as
nodes that are well-balanced in terms of how many other nodes are at various distances, for instance.\footnote{
Note that even the ordering produced by this class of measures is equivalent to ordering nodes according to $\sum_{\ell=1}^{L} \beta_{\ell} \log(\alpha_{\ell} + s_{i}^{\ell})$.  This is an additive form, with nodal statistics $\beta_{\ell} \log(\alpha_{\ell} + s_{i}^{\ell})$.
This shows that it can be challenging to escape the additive family.  Nonetheless, this is a new and potentially interesting family prompted by our analysis.}

\bibliographystyle{ecta}
\bibliography{networks2}

\section*{Appendix:  Proofs}

\noindent \textbf{Proof of Lemma \ref{psi}:} The if part is clear, and so we show the only if part.

Suppose that $s_i(\mathbf{g}) =
s_i(\mathbf{g}^{\prime })$ but $c_i(\mathbf{g}) \neq c_i(\mathbf{g}^{\prime })$. Without loss of
generality let $c_i(\mathbf{g})>c_i(\mathbf{g}^{\prime })$. By monotonicity, since $c_i(\mathbf{g})>c_i(\mathbf{g}^{\prime })$ it must be that $s_i(\mathbf{g}^{\prime }) \nsucceq
s_i(\mathbf{g})$.  However, this contradicts the fact that  $s_i(\mathbf{g}) = s_i(\mathbf{g}^{\prime })$
(which implies that $s_i(\mathbf{g}) \sim s_i(\mathbf{g}^{\prime })$ by reflexivity of a partial order)%
.
Thus, $c_i(\mathbf{g}) = c_i(\mathbf{g}^{\prime })$ for any $\mathbf{g},\mathbf{g}^{\prime }$ for which $s_i(\mathbf{g}) =
s_i(\mathbf{g}^{\prime })$.
Letting $S$ denote the range of $s_i(g)$ (which is the same for all $i$ by anonymity),
it follows that there exists  $\mathcal{C}_i:S\rightarrow \Re$
for which $c_i(\mathbf{g})=\mathcal{C}_i(s_i(\mathbf{g}))$ for any $\mathbf{g}$.  Moreover $\mathcal{C}_i$ must be a monotone function on $S$,
given the monotonicity of $c_i$.

Next, we show that $\mathcal{C}_i=\mathcal{C}_j$ for any $i,j$.
Consider any $s'\in S$ and any two nodes $i$ and $j$.  Since $s'\in S$, it follows that there exists $\mathbf{g}$ for which $%
s_i(\mathbf{g})=s'$.  Consider a permutation $\pi$ such that $\pi(j)=i$ and $%
\pi(i)=j$. Then by the anonymity of $s$, $s' =s_j(\mathbf{g}\circ \pi)$.  Thus, by anonymity of $c$, $c_i(\mathbf{g}) = c_j(\mathbf{g}\circ \pi)$ and so $\mathcal{C}_i(s')=\mathcal{C}_j(s')$.
 Given that $s'$ was arbitrary, it follows that $\mathcal{C}_i=\mathcal{C}_j=\mathcal{C}$ for some $\mathcal{C}:S\rightarrow \Re$ and all $i,j$.

We extend the function  $\mathcal{C}$ to be monotone on all of $\Re^L$ as follows.
Let $S_1$ be the set of $s\notin S$ such that
there exists some $s'\in S$ for which $s'\geq s$.   For any $s\in S_1$
$\mathcal{C}_i(s)=\inf_{s'\in S, s'\geq s} \mathcal{C}(s')$.
Next, let $S_2$ be the set of
 $s\notin S \cup S_1$.
For any $s\in S_2$ let $\mathcal{C}(s)=\sup_{s'\in S \cup S_1, s'\leq s} \mathcal{C}(s')$ (and note that this is well defined
for all $s\in S_2$ since there is always some $s'\in S \cup S_1, s'\leq s$) .
This is also monotone, by construction.\hfill $\Box$

\bigskip

\noindent \textbf{Proof of Theorems \ref{decay1}-\ref{decay3}}

\medskip \noindent \textbf{IF part:}

\noindent It is easily checked that if $\mathcal{C}$ can be expressed as in
equation (\ref{decay_eqn1}), then independence holds.
Similarly, if the representation is as in (\ref{decay_eqn2}), then recursivity also holds,
as does additivity if (\ref{decay_eqn3}) is satisfied.

\medskip \noindent \textbf{ONLY IF part:}

Let $\mathbf{e}^{\ell}$ denote the vector in $%
\mathbb{N}^{L}$ with every $\mathbf{e}_{\ell}^{\ell}=1$ and $\mathbf{e}_{j}^{\ell}=0$
for all $j\not=\ell$. Define $F_{\ell}:\mathbb{R}\rightarrow \mathbb{R}_{+}$ as%
\begin{equation}
F_{\ell}\left( x\right) =\mathcal{C}\left( x\mathbf{e}^{\ell}\right) \text{.}
\label{eq: define f_k}
\end{equation}%
Iterated applications of independence imply
that%
\begin{equation}
\mathcal{C}\left( s_{i}\right) =\sum_{\ell=1}^{L}F_{\ell}\left( s_{i,\ell}\right) .
\label{summability}
\end{equation}%
To see this, note that for $s_{i}=\left( x,0,...,0,...\right) $, $%
s_{i}^{\prime }=\left( 0,y,0,...,0,...\right) $, and $s_{i}^{\prime \prime
}=\left( x,y,0,...,0,...\right) $, independence requires%
\begin{eqnarray*}
&&\mathcal{C}\left( s_{i}^{\prime \prime }\right) -F_{2}\left( y\right)
=F_{1}\left( x\right) -0, \\
&&\mathcal{C}\left( s_{i}^{\prime \prime }\right) =F_{1}\left( x\right)
+F_{2}\left( y\right) .
\end{eqnarray*}%

Doing this again for $s_{i}=\left( x,y,0,0,...,0,...\right) $, $%
s_{i}^{\prime }=\left( x,0,z,0,...,0,...\right) $, and $s_{i}^{\prime \prime
}=\left( x,y,z,0,...,0,...\right) $, independence requires%
\begin{eqnarray*}
\mathcal{C}\left( s_{i}^{\prime \prime }\right) =F_{1}\left( x\right)
+F_{2}\left( y\right)+F_3(z).
\end{eqnarray*}
By induction, this holds for arbitrary vectors.
Monotonicity
implies that $F_{\ell}$ is increasing and $F_{\ell}\left( 0\right) =0$ for
all $\ell$.

By recursivity, for all $\ell\leq L$  and all $x$ in $\mathbb{R}$:%
\begin{equation}
\frac{F_{\ell+1}\left( x\right) }{F_{\ell}\left( x\right) }=\frac{F_{2}\left(
x\right) }{F_{1}\left( x\right) }=\delta \left( x\right)  \text{.}  \label{discount-x}
\end{equation}%
Moreover, recursivity also implies that for any two $x,x^{\prime }$
in $\Re$ and any $\ell$ (provided the denominators are not 0):%
\begin{equation}
\label{key1}
\frac{F_{\ell}\left( x^{\prime }\right) }{F_{\ell}\left( x\right) }=\frac{%
F_{1}\left( x^{\prime }\right) }{F_{1}\left( x\right) }.
\end{equation}%
From (\ref{discount-x}) this is true only if $\delta \left(
x\right) \equiv \delta $ is constant.

Next, note that (\ref{key1}), together with the fact $F_\ell(0)=0$ for all $\ell$,  imply that $F_\ell(x)=\delta^\ell f(x)$ for a common $f$ for which $f(0)=0$.
This implies that $\mathcal{C}
$ can be written as in equation (\ref{decay_eqn2}).

Finally, additivity -- which clearly implies independence -- then implies that $f$ is linear (a standard result in vector spaces), and given that it must be that $f(0)=0$, the final characterization follows.\eproof

\bigskip

\noindent \textbf{Proof of Theorems \ref{fulldecay} and \ref{degree}:}
The ``if'' parts of both theorems are straightforward, and so we prove the ``only if'' claims, beginning with
Theorem \ref{fulldecay}.

From Theorem \ref{decay3} it follows that
\begin{equation*}
c_{i}\left( g\right) =\mathcal{C}(s_{i}\left( g\right)
)=a \sum_{\ell=1}^{L}\delta ^{\ell-1}s_{i}^{\ell}\left( g\right).
\end{equation*}
The strictness of sensitivity implies that $\delta>0$ and $a>0$.
It suffices to show that $s_i(g)$ must be a weighted neighborhood statistic that is $\delta$-monotone.

Cycle independence implies that for any $g$,  $c_i(g) =  c_i(g')$ where $g'$ is any minimal $i$-centered subtree of $g$.  Thus, it suffices to prove the characterization for
trees.

Let us construct a series of line networks, with $i$ at one extreme and with $k$ links,  denoted $g^k$.
Iteratively, for $k=1$ to $n-1$ define
\[
w_i^k \equiv  \frac{\sum_{\ell=1}^{L}\delta ^{\ell-1}s_{i}^{\ell}\left( g^k\right) -   \sum_{\ell=1}^{L}\delta ^{\ell-1}s_{i}^{\ell}\left( g^{k-1}\right)}{\delta^\ell}.
\]

It follows directly (noting that $n_{i}^{\ell}\left( g^k\right)=1$ for $\ell\leq k$ and 0 otherwise) that
\begin{equation*}
c_{i}\left( g^k\right)=a \sum_{\ell=1}^{L}\delta ^{\ell-1} w_i^\ell n_{i}^{\ell}\left( g^k\right).
\end{equation*}

Any $i$-centered tree $g$ with depth (maximum distance to $i$) of $k$, can be built from $g^k$ by a successive addition of links.
Iteratively building $g$ from $g^k$ by adding links that connect to the tree present at each step, and applying the constant marginal values condition at each step, then implies that
\begin{equation*}
c_{i}\left( g\right)=a \sum_{\ell=1}^{L}\delta ^{\ell-1} w_i^\ell n_{i}^{\ell}\left( g\right).
\end{equation*}

Finally, comparing $c_i(g^k)$ to $c_i(g^{k-1}+ hj)$ where $h$ is at distance $k-2$ and $j$ is not in the network $g^{k-1}$, implies
that $\delta ^{\ell-1} w_i^\ell  > \delta ^{\ell } w_i^{\ell+1}$.

The proof  Theorem \ref{degree} comes from changing the final step above, which implies instead that  $\delta ^{\ell-1} w_i^\ell =0$ for all $\ell>1$.
\eproof

\bigskip

\noindent \textbf{Proof of Proposition \ref{theomonhier}:}

\medskip
\noindent [IF]
We first prove the `if' part.
Consider a monotone hierarchy and two nodes $i,j$.

First, suppose that $i \doteq j$. From the definition of $\doteq$ it must be that $i$ and $j$ are at the same
level of the hierarchy, and that any subgraph containing $i$ starting at the same level as some subgraph containing $j$ are identical (up to the labeling of the nodes).
Hence $i$ and $j$ are symmetric, and $n_i({\bf g})  = n_j({\bf g})$.

So, to complete the proof of the `if' part, it is sufficient to show
that if
$i \gtrdot j$ then $n_i \succ n_j$.

Consider two nodes at the same distance from the root.

If they are at distance 1, then they have the same distance to each node that is not a successor of either node.
Given the definition of monotone hierarchy, it must be that either $d_i>d_j$ or that $d_i=d_j$ and $d_k > d_l$ for some successor $k$ of $i$ and any successor $%
l $ of $j$ such that $\rho(k)= \rho(l)$.  In either case the definition implies that then $d_k \geq d_l$ for every successor $k$ of $i$ and successor $%
l $ of $j$ such that $\rho(k)= \rho(l)$.  it directly follows that $n_i \succ n_j$.

Inductively, if
$\rho(i) = \rho(j)>1$:
\begin{itemize}
\item If there exist two
distinct predecessors $k,l$ of $i,j$, respectively, such that $\rho(k) = \rho(l)$ and $%
k \gtrdot l$, then $i \gtrdot j$, then the ordering holds given the ordering of those predecessors and that their neighborhoods are determined by those predecessors.
\item Otherwise, they follow from a common immediate predecessor and differ only in the subgraphs starting from them,
and the condition follows from reasoning above given the differences in those subgraphs, which must be ordered.
\end{itemize}

Next, suppose that $%
\rho(j) = \rho(i)+1$.   We show that $n_i({\bf g}) \succ n_j({\bf g})$.

For this part of the proof we provide
a formula to compute the number of nodes at distance less than or equal to $%
d $ from node $i$ for a monotone hierarchy, $Q(i,d)$. Let $\rho(i)$ denote the distance from the root and $i+0,i_1,..,i_k,...,i_{\rho(i)}=i$  the unique path between the root and node $i$. Let $p(i,\ell)$
denote the number of successors of node $i$ at distance $\ell$.  If $d \geq \rho(i)$, we compute the number of nodes at distance less
than or equal to $d$ as

\begin{eqnarray*}
Q(i,d) & = & p(i_0,0) + p(i_0,1)+....+ p(i_0, d-\rho(i)) \\
& + & p(i_1,d-\rho(i)) + p(i_1, d-\rho(i)+1)+p(i_2, d-\rho(i)+1) + p(i_2, d-\rho(i)+2) \\
& + & ... p(i_{\rho(i)-1}, d-2) + p(i_{\rho(i)-1}, d-1) + p(i,d-1)+ p(i,d).
\end{eqnarray*}

To understand this computation, notice that all nodes which are at distance
less than or equal to $d-\rho(i)$ from the root are at a distance less than $d$
from node $i$. Other nodes at a distance less than $d$ from node $i$ are
computed considering the path between $i_0$ and $i$. Fix $i_1$. There are successor
nodes which are at distance $d-\rho(i)$ from node $i_1$ (and hence at a distance $d-1$ from $i$) and were not counted
earlier because they are at a distance of $d-\rho(i)+1$ from the root,  and
successor nodes which are at a distance $d-\rho(i)+1$ from node $1$ (and hence at a distance $d$ from $i$) and were not counted earlier because they are at a distance $d- \rho(i)+2$ from the root. Continuing along the
path, for any node $i_k$ we count successor nodes at a distance $d-\rho(i)+k-1$ and $d-\rho(i)+k$ from node $i_k$ which are at a distance $d-1$ and $d$ from node $i$ and were not counted earlier, and finally obtain the total number of nodes
at a distance less or equal to $d$ from node $i$.

Next suppose that $d \leq \rho(i)$. In that case, no node beyond $i_0$ who does
not belong to the subtree starting at $i_1$ can be at a distance smaller
than $d$. The expression for the number of nodes at a distance less than or
equal to $d$ simplifies to

\begin{eqnarray*}
Q(i,d) & = & p(i_{\rho(i)-d},0) + p(i_{\rho(i)-d+1},0)+ p(i_{\rho(i)-d+1}, 1) \\
& + & ...p(i_{\rho(i)-1},d-2)+ p(i_{\rho(i)-1},d-1)+ p(i,d-1) + p(i,d).
\end{eqnarray*}

\medskip

The following claim is useful.
\begin{claim}  \label{claimpath}In a monotone hierarchy, for any $i,j$ such that $\rho(j)=\rho(i)+1$, and any $\ell$, $p(i,\ell) \geq p(j,\ell)$.
\end{claim}

\bigskip

\noindent{\bf Proof of the Claim:} The proof is by induction on $\ell$. For $\ell=1$, the statement is true as $p(i,1) \equiv d_i-1 \geq d_j-1 \equiv p(j,1)$.
Suppose that the statement is true for all $\ell^{\prime}<\ell$. Let $i_1,..,i_I$ be the  direct successors of $i$ and $j_1,..,j_J$  the direct successors of $j$, with $J \leq I$. Then
\begin{eqnarray*}
 p(i, \ell)& = & \sum_{r=1}^I p(i_r,\ell-1), \\
 & \geq & \sum_{r=1}^J p(i_r,\ell-1), \\
 & \geq & \sum_{r=1}^J p(j_r,\ell-1) \\
 & = & p(j, \ell).
 \end{eqnarray*}

 \noindent where the first inequality is due to the fact that $I \geq J$ and the second that, by the induction hypothesis, as $\rho(i_r) = \rho(j_r)-1$ for all $r$, $p(i_r,\ell^{\prime}-1) \geq p(j_r, \ell^{\prime}-1)$.\eproof

Consider $d \geq \rho(i)+1$ and $i_0,i_1,.i_r,.,i_{\rho(i)}$, $%
i_0,j_1,...,j_r,  j_{\rho(i)+1}$ the paths linking $i$ and $j$ to
the root. Then
\begin{eqnarray*}
Q(i,d)-Q(j,d) & = & p(i_0, d- \rho(i))-p(j_1, d-\rho(i)-1) - p(j_1, d- \rho(i)) \\
& + & [p(i_1,d-\rho(i))+ p(i_1, d-\rho(i)+1)-p(j_2,d-\rho(i)) - p(j_2, d- \rho(i)+1) ]\\
& + & ... [p(i_r, d-\rho(i)+r-1) + p(i_r, d-\rho(i)+r)  \\
& & - p(j_{r+1}, d- \rho(i)+r-1)- p(j_{r+1}, d- \rho(i)+)] \\
&+ &... [p(i,d-1) +p(i,d)-p(j,d-1) -p(j,d)]
\end{eqnarray*}

Note that $p(i_0, d-\rho(i))= p(j_1, d-\rho(i)-1) + \sum_{k \neq j_1, \rho(k)=1} p(k, d- \rho(i)-1)$ and that $p(j_1, d- \rho(i)) = \sum_{l| \rho(l)=2, \rho(j_1,l)=1} p(l, d- \rho(i)-1)$. By Claim \ref{claimpath}, as $\rho(l)= \rho(k+1)$, $p(l, d- \rho(i)-1) \leq p(k, d- \rho(i)-1)$ and as $d(j_1) \leq d_(i_0)-1$, $ \sum_{k \neq j_1, \rho(k)=1} p(k, d- \rho(i)-1) \geq \sum_{l| \rho(l)=2, \rho(j_1,l)=1} p(l, d- \rho(i)-1)$. Furthermore, by Claim \ref{claimpath}, for all $r$ and all $d$, $p(i_r,d) \geq p(j_{r+1},d)$, so that $Q(i,d)-Q(j,d) \geq 0$.

Next, consider $d \leq \rho(i) < \rho(i)+1$. Then

\begin{eqnarray*}
Q(i,d)-Q(j,d) & = & [p(i_{\rho(i)-d},0) + p(i_{\rho(i)-d+1},0)-p(j_{\rho(i)-d+1},0) - p(j_{\rho(i)-d+2},0)] \\
& + & ... [p(i_{\rho(i)-1},d-1)+p(i,d-1) - p(j_{\rho(i)},d-1) - p(j,d-1)] \\
& + & [p(i,d)-p(j,d)]
\end{eqnarray*}
and by a direct application of Claim \ref{claimpath},  $Q(i,d)-Q(j,d) \geq 0$.

We finally observe
that there always exists a distance $d$ such that $Q(i,d)>Q(j,d)$. Let $h$ be the total number
of levels in the hierarchy. Consider a distance $d$ such that $h=d + \rho(i)$. Then
there exist successor nodes at distance $d$ from $i$ but no successor nodes at
distance $d$ from $j$. Hence $p(i,d)>0 = p(j,d)$. This establishes that $Q(i,d) > Q(j,d)$ and hence $n_i({\bf g}) \succ n_j({\bf g})$. By a repeated application of the same argument, for any $i,j$ such that $\rho(i) < \rho(j)$, for any $i,j$ such that $\rho(i,i_0) < \rho(j,i_0)$, $n_i({\bf g}) \succ n_j({\bf g})$.

\bigskip

[ONLY IF]: Suppose that the tree $g$ is not a monotone hierarchy and has an even diameter. Consider a line in the tree which has the same length as the diameter of the tree. Pick as a root the unique middle node in the line and let $h$ be the maximal distance between the root and a terminal node.

First consider the case in which there exist two nodes $i$ and $j$ such that $\rho(j)= \rho(i)+1$ but $d_j > d_i$. Then clearly $Q(j,1) > Q(k,1)$. Notice that all nodes are at a distance less than or equal to  $d=h+\rho(i)$ from node $i$ whereas there exist nodes which are at a distance $h+ \rho(i)+1$ from node $j$, and hence $Q(j,h+\rho(i)) < Q(i,h+ \rho(i))$ so that neither $n_i \succeq n_j$ nor $n_j \succeq n_i$.

Next suppose that for all nodes $i,j$ such that $\rho(j)=\rho(i)+1$, $d_j \leq d_i$, but that there exists two nodes $i, j$ at the same level of the hierarchy such that $d_i > d_j$ and two successors of $i$ and $j$, $k$ and $l$, at the same level of the hierarchy such that $d_k < d_l$. Because $d_i > d_j$, $Q(i,1)>Q(j,1)$. Suppose that $n_i \succ n_j$. Then $Q(i,d) > Q(j,d)$ fr all $d=1,2,...h+ \rho(i)-1$. Now consider the two successors $k$ and $l$ of $i$ and $j$. As $d_k < d_l$, $Q(k,1) < Q(l,1)$. Now count all the nodes which are at a distance less than  or equal to $h+\rho(k)-1$ from $k$, $Q(k,h+\rho(k)-1)$. This includes all the nodes but the nodes which are at maximal distance from $k$. As $k$ is a successor of $i$, the set of nodes at maximal distance from $k$ and $i$ are equal so that $Q(k,h+\rho(k)-1) =Q(i,h+\rho(i)-1)$. Similarly, the set of nodes at maximal distance from $j$ and $l$ are equal and $Q(l, h+ \rho(l)-1) = Q(j, h+ \rho(j)-1)$. Because we assume that $n_i \succ n_j$,  $Q(i,h+\rho(i)-1) > Q(j,h+\rho(j)-1) $ so that $Q(k, h + \rho(k)-1) > Q(j, h + \rho(j)-1)$, showing that neither $n_k \succ n_l$ nor $n_l \succ n_k$, completing the proof of the Proposition. \eproof

\bigskip

\noindent \textbf{Proof of Proposition \ref{theoregmonhier2}:} [IF] Because a
regular monotone hierarchy is a monotone hierarchy, we know by Proposition \ref%
{theomonhier} that  $i \gtrdot j$ if and only if $n_i \succ n_j$.

Let $d(\ell)$ be the degree of nodes at distance $\ell$ from the root node.

Next we show
that the number of geodesic paths of any length $d$ between two nodes is
smaller for a node further away from the root. To this end, consider two
nodes $i$ and $j$ such that $j$ is a direct successor of $i$. For any $d$,
if a geodesic path contains $j$ but not $i$, then $i$ must be an endpoint of
the path. Hence, the total number of geodesic paths of length $d$ going
through $j$ but not through $i$ is $2p(j,d-1)$. If $d_i \geq 3$, pick a
direct successor $k \neq j$ of $i$, and consider paths of length $d$
connecting successors of $k$ to $j$. All these paths must go through $i$ and
there are $2 p(k,d-1)= 2p(j,d-1)$ such paths. If $d_i=2$, then $d_j \leq 2$
so that $2p(j,d-1)=0$ or $2p(j,d-1)=2$. If $2p(j,d-1)=2$, then $d$ is small
enough so that there exists at least two paths of length $d$ connecting a
node in the network to $j$ through $i$. Furthermore, if $d = h-
\rho(i,i_0)+1 $ where $h$ is the number of levels of the hierarchy, there is
no path of length $d$ connecting $i$ to a node through $j$ whereas there
exist paths of length $d$ connecting a node to $j$ through $i$, so that $%
I_i \geq I_j$.

Next, we compute the number of walks emanating from two nodes $i$ and $j$ at
different levels of the hierarchy. Let $w_k(d)$ denote the number of walks
of length $d$ emanating from a node at level $\ell$. We show that $w_\ell(d) \geq
w_{\ell+1}(d)$. We compute the number of walks recursively:
\begin{eqnarray*}
w_\ell(d) & = & [d(\ell)-1] w_{\ell+1}(d-1) + w_{\ell-1}(d-1) \mbox{  for  } \ell \geq 1 \\
w_0(d) & = & d(0) w_1(d-1)
\end{eqnarray*}

We also have $w_\ell(0) = 1$ for all $\ell$ which allows us to start the recursion.

\bigskip

Next we prove that $w_d(\ell) \geq w_{\ell+1}(d)$ for $i=1,..,I-1$by induction on $%
d$ The statement is trivially true for all $\ell$ at $d=0$. Now suppose that
the statement is true at $d-1$. We first show that the inequality holds for
all nodes but the root. For $\ell \geq 1$,

\begin{eqnarray*}
w_{\ell}(d) & = & [d(\ell)-1] w_{\ell+1}(d-1) + w_{\ell-1}(d-1) \\
& \geq & [d(\ell+1)-1]w_{\ell+2}(d-1) + w_{\ell}(d-1) \\
& = & w_{\ell+1}(d)
\end{eqnarray*}

\noindent The more difficult step is to show that the statement is also true
for the root. To this end, we prove by induction on $d$ that for all $%
\ell=1,..,h-1$:

\begin{equation*}
d(0) w_\ell(d) \geq [d(\ell)-1] w_{\ell+1}(d) + w_{\ell-1}(d),
\end{equation*}

The statement is true at $d=0$ because $d(0)\geq d(\ell)$ for all $\ell \geq 1$.
Next compute

\begin{eqnarray*}
d(0) w_\ell(d) & = & d(0)[[d(\ell)-1]w_{\ell+1}(d-1) + w_{\ell-1}(d-1)], \\
(d(\ell-1)-1) w_{\ell+1}(d) + w_{\ell-1}(d) & = & [d(\ell)-1] [[d(\ell+2)-1]w_{\ell+2}(d-1) +
w_\ell(d-1)] \\
& + & [d(\ell-1)-1] w_\ell(d-1) + w_{\ell-2}(d-1).
\end{eqnarray*}

By the induction hypothesis,

\begin{equation*}
d(0) w_{\ell-1}(d-1) \geq [d(\ell-1)-1] w_\ell(d-1) + w_{\ell-2}(d-1),
\end{equation*}

\noindent and

\begin{equation*}
d(0) w_{\ell+1}(d-1) \geq [d(\ell+1)-1] w_{\ell+2}(d-1) + w_\ell(d-1) \geq [d(\ell+2)-1]
w_{\ell+2}(d-1) + w_\ell(d-1).
\end{equation*}

Replacing, we obtain

\begin{equation*}
d(0) w_\ell(d) \geq [d(\ell)-1] w_{\ell+1}(d) + w_{\ell-1}(d),
\end{equation*}

\noindent concluding the inductive argument. Applying this formula for $\ell=1$%
, we have $w_0(d) = d(0) w_1(d-1) \geq [d(1)-1] w_2(d-1) + w_0(d-1) = w_1(d)$%
, completing the proof that $w_\ell(d) \geq w_{\ell+1}(d)$ for all $d$.

\bigskip

[ONLY IF] Consider a leaf $i$ of the tree. Then $w_i=(0,0...0)$. So all
leaves have the same centrality based on the intermediary statistic. They must
also have the same centrality based on the neighborhood statistic, which implies
that the tree is a regular monotone hierarchy.\eproof

\newpage

\section*{Appendix:  Simulations -- Differences in Centrality Measures by Network Type}

We simulate networks on 40 nodes.
We vary the type of network to have three different basic structures, corresponding to a standard random network, a simple version of a stochastic block model, and a variation of a stochastic block model that includes bridge nodes.
The first is an Erdos-Renyi random graph in which all links are formed independently.
The second is a network that has some homophily: there are two types of nodes and we connect nodes of the same type with a different probability than nodes of different types.
The third is a variant of a homophilous network in which some nodes are `bridge nodes' that connect to other nodes with a uniform probability, thus putting them as connector nodes between the two homophilitic groups.
We vary the overall average degrees of the networks to be either 2, 5 or 10.
In the cases of the homophily and homophily bridge nodes, there are also relative within and across group link probabilities that vary.
Given all of these dimensions, we end up with many different networks on which to compare centrality measures.

We then compare 5 different centrality measures on these networks:  degree, decay, closeness, diffusion, and Katz-Bonacich.
Decay, diffusion and Katz-Bonacich all depend on a parameter that we call the exponential parameter, and we vary that as well.\footnote{In addition, diffusion centrality has $T=5$ in all of the simulations.}

The details on the three network types we perform the simulation on are:
\begin{itemize}
\item

\textbf{ER random graphs:} Each possible  link is formed independently with probability
$p=\overline{d}/\left( n-1\right) $.

\item  \textbf{Homophily:}

There are two equally-sized groups of $20$ nodes.  Links between pairs of nodes in the same group are formed with probability
$p_{same}$ and between pairs of nodes in different groups are formed with probability $p_{diff}$, all independently.

Letting $p_{same}=H\times p_{diff}$, average degree is:%
\[
\overline{d} =\left( \frac{n}{2}-1\right) p_{same}+\frac{n}{2}p_{diff}
\]

\item \textbf{Homophily with Bridge Nodes:}

There are
$L$ bridge nodes  and two equal-sized groups of $\frac{nN-L}{2}$ non-bridge nodes.
Each bridge node connects to any other node with probability $p_{b}$.
Non-bridge nodes connect to other same group nodes with probability $p_{same}$ and different group nodes with probability $p_{diff}$.

Letting $p_{same}=H\times p_{diff}$,  we set
$
p_{b} =\overline{d}/\left( n-1\right)$  where $\overline{d}$ is defined as the average degree
\[
\overline{d} =\left( \frac{n-L}{2}-1\right) p_{same}+\left( \frac{n-L}{2}%
\right) p_{diff}+Lp_{b}.
\]
\end{itemize}

\medskip

A first thing to note about the simulations (see the tables below) is that the correlation in rankings of the various centrality measures is very high across all of the simulations and measures, often above .9,  and usually in the .8 to 1 range.
This is in part reflective of what we have seen from our characterizations:  all of these measures operate in a similar manner and are based on nodal statistics that often move in similar ways:  nodes with higher degree tend to be closer to other nodes and have more walks to other nodes, and so forth.
In terms of differences between measures,  closeness and betweenness are more distinguished from the others in terms of correlation, while the other measures all correlate above .98 in Table \ref{tab:er1}.

These extreme correlations are higher than those found in
\citet*{valente2008}, who also find high correlations, but lower in magnitude, when looking at a series of real data sets.\footnote{See \cite{schochvb2017} for some discussion of
how correlation varies with network structure. }
The artificial nature of the Erdos-Renyi networks serves as a benchmark from which we can jump off as it results in less differentiation between nodes than one finds in many real-world networks, but also allows us to know that differences among nodes are coming from random variations.  As we add homophily in Table \ref{tab:hom1}, and then bridge nodes in Table \ref{tab:hombridge1}, we see the correlations drop significantly, especially comparing betweenness centrality to the others, and this then has an intuitive interpretation as bridge nodes naturally have high betweenness centrality, but may not stand out according to other measures.

Correlation is a very crude measure, and it does not capture whether nodes are switching ranking or by how much.  Some nodes could have dramatically different rankings and yet the correlation could be relatively high overall.
Thus, we also look at how many nodes switch rankings between two measures, as well as how the maximal extent to which some node changes rankings.
There, we see
more substantial differences across centrality measures, and with most measures being more highly distinguished from each other.

As we increase the exponential factor (e.g., from Table \ref{tab:er1} to \ref{tab:er2}), we see greater differences between the measures, as the correlations drop and we see more changes in the rankings.   With a very low exponential factor, decay, diffusion, Katz-Bonacich are all very close to degree, while for higher exponential parameters they begin to differentiate themselves.   This makes sense as it allows the measures to incorporate information that depends on more of the network, and that is less tied to immediate neighborhoods.

\newpage


\begin{table}[htbp]
  \centering
  \caption{Avg degree 2,  Erdos Renyi,  Decay .15}
    \begin{tabular}{llcccc}
    \toprule
 &    & Correlation & Fraction of Sims & Fraction Nodes & Max Change \\
 Cent1 & Cent2 &  of Rank Vector & w same Top Node & Switch Rank & in \% Rank \\
    \midrule
   Degree	&	Decay	&	0.99	&	1.00	&	0.73	&	0.16	\\	
Degree	&	Closeness	&	0.83	&	0.78	&	0.77	&	0.34	\\	
Degree	&	Diffusion	&	0.98	&	0.98	&	0.74	&	0.17	\\	
Degree	&	KatzBon	&	1.00	&	1.00	&	0.73	&	0.16	\\	
Degree	&	Between	&	0.83	&	0.50	&	0.93	&	0.30	\\	\midrule
Decay	&	Degree	&	0.99	&	1.00	&	0.73	&	0.16	\\	
Decay	&	Closeness	&	0.88	&	0.74	&	0.61	&	0.23	\\	
Decay	&	Diffusion	&	1.00	&	0.86	&	0.30	&	0.08	\\	
Decay	&	KatzBon	&	0.99	&	0.86	&	0.29	&	0.08	\\	
Decay	&	Between	&	0.84	&	0.42	&	0.90	&	0.32	\\	\midrule
Closeness	&	Degree	&	0.83	&	0.78	&	0.77	&	0.34	\\	
Closeness	&	Decay	&	0.88	&	0.74	&	0.61	&	0.23	\\	
Closeness	&	Diffusion	&	0.86	&	0.66	&	0.65	&	0.24	\\	
Closeness	&	KatzBon	&	0.83	&	0.66	&	0.65	&	0.25	\\	
Closeness	&	Between	&	0.67	&	0.52	&	0.91	&	0.35	\\	\midrule
Diffusion	&	Degree	&	0.98	&	0.98	&	0.74	&	0.17	\\	
Diffusion	&	Decay	&	1.00	&	0.86	&	0.30	&	0.08	\\	
Diffusion	&	Closeness	&	0.86	&	0.66	&	0.65	&	0.24	\\	
Diffusion	&	KatzBon	&	0.99	&	0.98	&	0.11	&	0.04	\\	
Diffusion	&	Between	&	0.83	&	0.34	&	0.91	&	0.36	\\	\midrule
KatzBon	&	Degree	&	1.00	&	1.00	&	0.73	&	0.16	\\	
KatzBon	&	Decay	&	0.99	&	0.86	&	0.29	&	0.08	\\	
KatzBon	&	Closeness	&	0.83	&	0.66	&	0.65	&	0.25	\\	
KatzBon	&	Diffusion	&	0.99	&	0.98	&	0.11	&	0.04	\\	
KatzBon	&	Between	&	0.83	&	0.34	&	0.91	&	0.35	\\	\midrule
Between	&	Degree	&	0.83	&	0.50	&	0.93	&	0.30	\\	
Between	&	Decay	&	0.84	&	0.42	&	0.90	&	0.32	\\	
Between	&	Closeness	&	0.67	&	0.52	&	0.91	&	0.35	\\	
Between	&	Diffusion	&	0.83	&	0.34	&	0.91	&	0.36	\\	
Between	&	KatzBon	&	0.83	&	0.34	&	0.91	&	0.35	\\	

    \bottomrule
    \end{tabular}%
  \label{tab:er1}%
\end{table}%

\begin{table}[htbp]
  \centering
  \caption{Avg degree 2,  Erdos Renyi,  Decay .5}
    \begin{tabular}{llcccc}
    \toprule
 &    & Correlation & Fraction of Sims & Fraction Nodes & Max Change \\
 Cent1 & Cent2 &  of Rank Vector & w same Top Node & Switch Rank & in \% Rank \\
    \midrule
  Degree	&	Decay	&	0.89	&	0.80	&	0.78	&	0.34	\\	
Degree	&	Closeness	&	0.82	&	0.84	&	0.78	&	0.33	\\	
Degree	&	Diffusion	&	0.91	&	0.96	&	0.78	&	0.34	\\	
Degree	&	KatzBon	&	1.00	&	1.00	&	0.74	&	0.17	\\	
Degree	&	Between	&	0.84	&	0.66	&	0.92	&	0.31	\\	\midrule
Decay	&	Degree	&	0.89	&	0.80	&	0.78	&	0.34	\\	
Decay	&	Closeness	&	0.97	&	0.96	&	0.26	&	0.07	\\	
Decay	&	Diffusion	&	0.93	&	0.68	&	0.57	&	0.17	\\	
Decay	&	KatzBon	&	0.90	&	0.66	&	0.65	&	0.25	\\	
Decay	&	Between	&	0.78	&	0.78	&	0.91	&	0.39	\\	\midrule
Closeness	&	Degree	&	0.82	&	0.84	&	0.78	&	0.33	\\	
Closeness	&	Decay	&	0.97	&	0.96	&	0.26	&	0.07	\\	
Closeness	&	Diffusion	&	0.84	&	0.70	&	0.61	&	0.19	\\	
Closeness	&	KatzBon	&	0.83	&	0.70	&	0.66	&	0.24	\\	
Closeness	&	Between	&	0.67	&	0.80	&	0.91	&	0.38	\\	\midrule
Diffusion	&	Degree	&	0.91	&	0.96	&	0.78	&	0.34	\\	
Diffusion	&	Decay	&	0.93	&	0.68	&	0.57	&	0.17	\\	
Diffusion	&	Closeness	&	0.84	&	0.70	&	0.61	&	0.19	\\	
Diffusion	&	KatzBon	&	0.92	&	0.92	&	0.58	&	0.23	\\	
Diffusion	&	Between	&	0.80	&	0.56	&	0.93	&	0.46	\\	\midrule
KatzBon	&	Degree	&	1.00	&	1.00	&	0.74	&	0.17	\\	
KatzBon	&	Decay	&	0.90	&	0.66	&	0.65	&	0.25	\\	
KatzBon	&	Closeness	&	0.83	&	0.70	&	0.66	&	0.24	\\	
KatzBon	&	Diffusion	&	0.92	&	0.92	&	0.58	&	0.23	\\	
KatzBon	&	Between	&	0.84	&	0.54	&	0.91	&	0.35	\\	\midrule
Between	&	Degree	&	0.84	&	0.66	&	0.92	&	0.31	\\	
Between	&	Decay	&	0.78	&	0.78	&	0.91	&	0.39	\\	
Between	&	Closeness	&	0.67	&	0.80	&	0.91	&	0.38	\\	
Between	&	Diffusion	&	0.80	&	0.56	&	0.93	&	0.46	\\	
Between	&	KatzBon	&	0.84	&	0.54	&	0.91	&	0.35	\\	

    \bottomrule
    \end{tabular}%
  \label{tab:er2}%
\end{table}%

\begin{table}[htbp]
  \centering
  \caption{Avg degree 2,  Homophily,  Decay .5}
    \begin{tabular}{llcccc}
    \toprule
 &    & Correlation & Fraction of Sims & Fraction Nodes & Max Change \\
 Cent1 & Cent2 &  of Rank Vector & w same Top Node & Switch Rank & in \% Rank \\
    \midrule
  Degree	&	Decay	&	0.89	&	0.80	&	0.80	&	0.35	\\	
Degree	&	Closeness	&	0.82	&	0.84	&	0.79	&	0.37	\\	
Degree	&	Diffusion	&	0.90	&	0.88	&	0.79	&	0.34	\\	
Degree	&	KatzBon	&	1.00	&	1.00	&	0.75	&	0.15	\\	
Degree	&	Between	&	0.79	&	0.54	&	0.93	&	0.34	\\	\midrule
Decay	&	Degree	&	0.89	&	0.80	&	0.80	&	0.35	\\	
Decay	&	Closeness	&	0.97	&	0.96	&	0.33	&	0.07	\\	
Decay	&	Diffusion	&	0.92	&	0.64	&	0.62	&	0.18	\\	
Decay	&	KatzBon	&	0.90	&	0.68	&	0.69	&	0.26	\\	
Decay	&	Between	&	0.75	&	0.64	&	0.91	&	0.40	\\	\midrule
Closeness	&	Degree	&	0.82	&	0.84	&	0.79	&	0.37	\\	
Closeness	&	Decay	&	0.97	&	0.96	&	0.33	&	0.07	\\	
Closeness	&	Diffusion	&	0.83	&	0.64	&	0.64	&	0.19	\\	
Closeness	&	KatzBon	&	0.83	&	0.70	&	0.68	&	0.28	\\	
Closeness	&	Between	&	0.65	&	0.60	&	0.92	&	0.39	\\	\midrule
Diffusion	&	Degree	&	0.90	&	0.88	&	0.79	&	0.34	\\	
Diffusion	&	Decay	&	0.92	&	0.64	&	0.62	&	0.18	\\	
Diffusion	&	Closeness	&	0.83	&	0.64	&	0.64	&	0.19	\\	
Diffusion	&	KatzBon	&	0.91	&	0.80	&	0.63	&	0.24	\\	
Diffusion	&	Between	&	0.75	&	0.38	&	0.93	&	0.47	\\	\midrule
KatzBon	&	Degree	&	1.00	&	1.00	&	0.75	&	0.15	\\	
KatzBon	&	Decay	&	0.90	&	0.68	&	0.69	&	0.26	\\	
KatzBon	&	Closeness	&	0.83	&	0.70	&	0.68	&	0.28	\\	
KatzBon	&	Diffusion	&	0.91	&	0.80	&	0.63	&	0.24	\\	
KatzBon	&	Between	&	0.80	&	0.44	&	0.91	&	0.36	\\	\midrule
Between	&	Degree	&	0.79	&	0.54	&	0.93	&	0.34	\\	
Between	&	Decay	&	0.75	&	0.64	&	0.91	&	0.40	\\	
Between	&	Closeness	&	0.65	&	0.60	&	0.92	&	0.39	\\	
Between	&	Diffusion	&	0.75	&	0.38	&	0.93	&	0.47	\\	
Between	&	KatzBon	&	0.80	&	0.44	&	0.91	&	0.36	\\	

    \bottomrule
    \end{tabular}%
  \label{tab:hom1}%
\end{table}%

\begin{table}[htbp]
  \centering
  \caption{Avg degree 2,  Homophily-Bridge,  Decay .5}
    \begin{tabular}{llcccc}
    \toprule
 &    & Correlation & Fraction of Sims & Fraction Nodes & Max Change \\
 Cent1 & Cent2 &  of Rank Vector & w same Top Node & Switch Rank & in \% Rank \\
    \midrule
Degree	&	Decay	&	0.89	&	0.82	&	0.80	&	0.34	\\	
Degree	&	Closeness	&	0.82	&	0.86	&	0.79	&	0.36	\\	
Degree	&	Diffusion	&	0.90	&	0.88	&	0.79	&	0.34	\\	
Degree	&	KatzBon	&	1.00	&	1.00	&	0.75	&	0.15	\\	
Degree	&	Between	&	0.79	&	0.56	&	0.93	&	0.34	\\	\midrule
Decay	&	Degree	&	0.89	&	0.82	&	0.80	&	0.34	\\	
Decay	&	Closeness	&	0.97	&	0.94	&	0.32	&	0.07	\\	
Decay	&	Diffusion	&	0.93	&	0.66	&	0.61	&	0.18	\\	
Decay	&	KatzBon	&	0.90	&	0.70	&	0.68	&	0.25	\\	
Decay	&	Between	&	0.75	&	0.58	&	0.92	&	0.40	\\	\midrule
Closeness	&	Degree	&	0.82	&	0.86	&	0.79	&	0.36	\\	
Closeness	&	Decay	&	0.97	&	0.94	&	0.32	&	0.07	\\	
Closeness	&	Diffusion	&	0.84	&	0.66	&	0.63	&	0.19	\\	
Closeness	&	KatzBon	&	0.83	&	0.72	&	0.67	&	0.27	\\	
Closeness	&	Between	&	0.65	&	0.58	&	0.92	&	0.39	\\	\midrule
Diffusion	&	Degree	&	0.90	&	0.88	&	0.79	&	0.34	\\	
Diffusion	&	Decay	&	0.93	&	0.66	&	0.61	&	0.18	\\	
Diffusion	&	Closeness	&	0.84	&	0.66	&	0.63	&	0.19	\\	
Diffusion	&	KatzBon	&	0.91	&	0.80	&	0.62	&	0.24	\\	
Diffusion	&	Between	&	0.75	&	0.40	&	0.93	&	0.46	\\	\midrule
KatzBon	&	Degree	&	1.00	&	1.00	&	0.75	&	0.15	\\	
KatzBon	&	Decay	&	0.90	&	0.70	&	0.68	&	0.25	\\	
KatzBon	&	Closeness	&	0.83	&	0.72	&	0.67	&	0.27	\\	
KatzBon	&	Diffusion	&	0.91	&	0.80	&	0.62	&	0.24	\\	
KatzBon	&	Between	&	0.79	&	0.46	&	0.91	&	0.36	\\	\midrule
Between	&	Degree	&	0.79	&	0.56	&	0.93	&	0.34	\\	
Between	&	Decay	&	0.75	&	0.58	&	0.92	&	0.40	\\	
Between	&	Closeness	&	0.65	&	0.58	&	0.92	&	0.39	\\	
Between	&	Diffusion	&	0.75	&	0.40	&	0.93	&	0.46	\\	
Between	&	KatzBon	&	0.79	&	0.46	&	0.91	&	0.36	\\

    \bottomrule
    \end{tabular}%
  \label{tab:hombridge1}%
\end{table}%

\end{document}